\documentclass[11pt]{article}
\usepackage{amsmath,amsfonts,sint,epsfig,cite,graphics}
\usepackage{amssymb,bbm}
\usepackage{mathrsfs}  
\usepackage{amsbsy,color}
\usepackage{latexsym}
\usepackage{graphicx}
\usepackage{psfrag}
\usepackage{macros_static}
\usepackage{url}
\definecolor{AliceBlue}{rgb}{0.94,0.97,1.00}
\definecolor{AntiqueWhite1}{rgb}{1.00,0.94,0.86}
\definecolor{AntiqueWhite2}{rgb}{0.93,0.87,0.80}
\definecolor{AntiqueWhite3}{rgb}{0.80,0.75,0.69}
\definecolor{AntiqueWhite4}{rgb}{0.55,0.51,0.47}
\definecolor{AntiqueWhite}{rgb}{0.98,0.92,0.84}
\definecolor{BlanchedAlmond}{rgb}{1.00,0.92,0.80}
\definecolor{BlueViolet}{rgb}{0.54,0.17,0.89}
\definecolor{CadetBlue1}{rgb}{0.60,0.96,1.00}
\definecolor{CadetBlue2}{rgb}{0.56,0.90,0.93}
\definecolor{CadetBlue3}{rgb}{0.48,0.77,0.80}
\definecolor{CadetBlue4}{rgb}{0.33,0.53,0.55}
\definecolor{CadetBlue}{rgb}{0.37,0.62,0.63}
\definecolor{CornflowerBlue}{rgb}{0.39,0.58,0.93}
\definecolor{DarkBlue}{rgb}{0.00,0.00,0.55}
\definecolor{DarkCyan}{rgb}{0.00,0.55,0.55}
\definecolor{DarkGoldenrod1}{rgb}{1.00,0.73,0.06}
\definecolor{DarkGoldenrod2}{rgb}{0.93,0.68,0.05}
\definecolor{DarkGoldenrod3}{rgb}{0.80,0.58,0.05}
\definecolor{DarkGoldenrod4}{rgb}{0.55,0.40,0.03}
\definecolor{DarkGoldenrod}{rgb}{0.72,0.53,0.04}
\definecolor{DarkGray}{rgb}{0.66,0.66,0.66}
\definecolor{DarkGreen}{rgb}{0.00,0.39,0.00}
\definecolor{DarkGrey}{rgb}{0.66,0.66,0.66}
\definecolor{DarkKhaki}{rgb}{0.74,0.72,0.42}
\definecolor{DarkMagenta}{rgb}{0.55,0.00,0.55}
\definecolor{DarkOliveGreen1}{rgb}{0.79,1.00,0.44}
\definecolor{DarkOliveGreen2}{rgb}{0.74,0.93,0.41}
\definecolor{DarkOliveGreen3}{rgb}{0.64,0.80,0.35}
\definecolor{DarkOliveGreen4}{rgb}{0.43,0.55,0.24}
\definecolor{DarkOliveGreen}{rgb}{0.33,0.42,0.18}
\definecolor{DarkOrange1}{rgb}{1.00,0.50,0.00}
\definecolor{DarkOrange2}{rgb}{0.93,0.46,0.00}
\definecolor{DarkOrange3}{rgb}{0.80,0.40,0.00}
\definecolor{DarkOrange4}{rgb}{0.55,0.27,0.00}
\definecolor{DarkOrange}{rgb}{1.00,0.55,0.00}
\definecolor{DarkOrchid1}{rgb}{0.75,0.24,1.00}
\definecolor{DarkOrchid2}{rgb}{0.70,0.23,0.93}
\definecolor{DarkOrchid3}{rgb}{0.60,0.20,0.80}
\definecolor{DarkOrchid4}{rgb}{0.41,0.13,0.55}
\definecolor{DarkOrchid}{rgb}{0.60,0.20,0.80}
\definecolor{DarkRed}{rgb}{0.55,0.00,0.00}
\definecolor{DarkSalmon}{rgb}{0.91,0.59,0.48}
\definecolor{DarkSeaGreen1}{rgb}{0.76,1.00,0.76}
\definecolor{DarkSeaGreen2}{rgb}{0.71,0.93,0.71}
\definecolor{DarkSeaGreen3}{rgb}{0.61,0.80,0.61}
\definecolor{DarkSeaGreen4}{rgb}{0.41,0.55,0.41}
\definecolor{DarkSeaGreen}{rgb}{0.56,0.74,0.56}
\definecolor{DarkSlateBlue}{rgb}{0.28,0.24,0.55}
\definecolor{DarkSlateGray1}{rgb}{0.59,1.00,1.00}
\definecolor{DarkSlateGray2}{rgb}{0.55,0.93,0.93}
\definecolor{DarkSlateGray3}{rgb}{0.47,0.80,0.80}
\definecolor{DarkSlateGray4}{rgb}{0.32,0.55,0.55}
\definecolor{DarkSlateGray}{rgb}{0.18,0.31,0.31}
\definecolor{DarkSlateGrey}{rgb}{0.18,0.31,0.31}
\definecolor{DarkTurquoise}{rgb}{0.00,0.81,0.82}
\definecolor{DarkViolet}{rgb}{0.58,0.00,0.83}
\definecolor{DeepPink1}{rgb}{1.00,0.08,0.58}
\definecolor{DeepPink2}{rgb}{0.93,0.07,0.54}
\definecolor{DeepPink3}{rgb}{0.80,0.06,0.46}
\definecolor{DeepPink4}{rgb}{0.55,0.04,0.31}
\definecolor{DeepPink}{rgb}{1.00,0.08,0.58}
\definecolor{DeepSkyBlue1}{rgb}{0.00,0.75,1.00}
\definecolor{DeepSkyBlue2}{rgb}{0.00,0.70,0.93}
\definecolor{DeepSkyBlue3}{rgb}{0.00,0.60,0.80}
\definecolor{DeepSkyBlue4}{rgb}{0.00,0.41,0.55}
\definecolor{DeepSkyBlue}{rgb}{0.00,0.75,1.00}
\definecolor{DimGray}{rgb}{0.41,0.41,0.41}
\definecolor{DimGrey}{rgb}{0.41,0.41,0.41}
\definecolor{DodgerBlue1}{rgb}{0.12,0.56,1.00}
\definecolor{DodgerBlue2}{rgb}{0.11,0.53,0.93}
\definecolor{DodgerBlue3}{rgb}{0.09,0.45,0.80}
\definecolor{DodgerBlue4}{rgb}{0.06,0.31,0.55}
\definecolor{DodgerBlue}{rgb}{0.12,0.56,1.00}
\definecolor{FloralWhite}{rgb}{1.00,0.98,0.94}
\definecolor{ForestGreen}{rgb}{0.13,0.55,0.13}
\definecolor{GhostWhite}{rgb}{0.97,0.97,1.00}
\definecolor{GreenYellow}{rgb}{0.68,1.00,0.18}
\definecolor{HotPink1}{rgb}{1.00,0.43,0.71}
\definecolor{HotPink2}{rgb}{0.93,0.42,0.65}
\definecolor{HotPink3}{rgb}{0.80,0.38,0.56}
\definecolor{HotPink4}{rgb}{0.55,0.23,0.38}
\definecolor{HotPink}{rgb}{1.00,0.41,0.71}
\definecolor{IndianRed1}{rgb}{1.00,0.42,0.42}
\definecolor{IndianRed2}{rgb}{0.93,0.39,0.39}
\definecolor{IndianRed3}{rgb}{0.80,0.33,0.33}
\definecolor{IndianRed4}{rgb}{0.55,0.23,0.23}
\definecolor{IndianRed}{rgb}{0.80,0.36,0.36}
\definecolor{LavenderBlush1}{rgb}{1.00,0.94,0.96}
\definecolor{LavenderBlush2}{rgb}{0.93,0.88,0.90}
\definecolor{LavenderBlush3}{rgb}{0.80,0.76,0.77}
\definecolor{LavenderBlush4}{rgb}{0.55,0.51,0.53}
\definecolor{LavenderBlush}{rgb}{1.00,0.94,0.96}
\definecolor{LawnGreen}{rgb}{0.49,0.99,0.00}
\definecolor{LemonChiffon1}{rgb}{1.00,0.98,0.80}
\definecolor{LemonChiffon2}{rgb}{0.93,0.91,0.75}
\definecolor{LemonChiffon3}{rgb}{0.80,0.79,0.65}
\definecolor{LemonChiffon4}{rgb}{0.55,0.54,0.44}
\definecolor{LemonChiffon}{rgb}{1.00,0.98,0.80}
\definecolor{LightBlue1}{rgb}{0.75,0.94,1.00}
\definecolor{LightBlue2}{rgb}{0.70,0.87,0.93}
\definecolor{LightBlue3}{rgb}{0.60,0.75,0.80}
\definecolor{LightBlue4}{rgb}{0.41,0.51,0.55}
\definecolor{LightBlue}{rgb}{0.68,0.85,0.90}
\definecolor{LightCoral}{rgb}{0.94,0.50,0.50}
\definecolor{LightCyan1}{rgb}{0.88,1.00,1.00}
\definecolor{LightCyan2}{rgb}{0.82,0.93,0.93}
\definecolor{LightCyan3}{rgb}{0.71,0.80,0.80}
\definecolor{LightCyan4}{rgb}{0.48,0.55,0.55}
\definecolor{LightCyan}{rgb}{0.88,1.00,1.00}
\definecolor{LightGoldenrod1}{rgb}{1.00,0.93,0.55}
\definecolor{LightGoldenrod2}{rgb}{0.93,0.86,0.51}
\definecolor{LightGoldenrod3}{rgb}{0.80,0.75,0.44}
\definecolor{LightGoldenrod4}{rgb}{0.55,0.51,0.30}
\definecolor{LightGoldenrodYellow}{rgb}{0.98,0.98,0.82}
\definecolor{LightGoldenrod}{rgb}{0.93,0.87,0.51}
\definecolor{LightGray}{rgb}{0.83,0.83,0.83}
\definecolor{LightGreen}{rgb}{0.56,0.93,0.56}
\definecolor{LightGrey}{rgb}{0.83,0.83,0.83}
\definecolor{LightPink1}{rgb}{1.00,0.68,0.73}
\definecolor{LightPink2}{rgb}{0.93,0.64,0.68}
\definecolor{LightPink3}{rgb}{0.80,0.55,0.58}
\definecolor{LightPink4}{rgb}{0.55,0.37,0.40}
\definecolor{LightPink}{rgb}{1.00,0.71,0.76}
\definecolor{LightSalmon1}{rgb}{1.00,0.63,0.48}
\definecolor{LightSalmon2}{rgb}{0.93,0.58,0.45}
\definecolor{LightSalmon3}{rgb}{0.80,0.51,0.38}
\definecolor{LightSalmon4}{rgb}{0.55,0.34,0.26}
\definecolor{LightSalmon}{rgb}{1.00,0.63,0.48}
\definecolor{LightSeaGreen}{rgb}{0.13,0.70,0.67}
\definecolor{LightSkyBlue1}{rgb}{0.69,0.89,1.00}
\definecolor{LightSkyBlue2}{rgb}{0.64,0.83,0.93}
\definecolor{LightSkyBlue3}{rgb}{0.55,0.71,0.80}
\definecolor{LightSkyBlue4}{rgb}{0.38,0.48,0.55}
\definecolor{LightSkyBlue}{rgb}{0.53,0.81,0.98}
\definecolor{LightSlateBlue}{rgb}{0.52,0.44,1.00}
\definecolor{LightSlateGray}{rgb}{0.47,0.53,0.60}
\definecolor{LightSlateGrey}{rgb}{0.47,0.53,0.60}
\definecolor{LightSteelBlue1}{rgb}{0.79,0.88,1.00}
\definecolor{LightSteelBlue2}{rgb}{0.74,0.82,0.93}
\definecolor{LightSteelBlue3}{rgb}{0.64,0.71,0.80}
\definecolor{LightSteelBlue4}{rgb}{0.43,0.48,0.55}
\definecolor{LightSteelBlue}{rgb}{0.69,0.77,0.87}
\definecolor{LightYellow1}{rgb}{1.00,1.00,0.88}
\definecolor{LightYellow2}{rgb}{0.93,0.93,0.82}
\definecolor{LightYellow3}{rgb}{0.80,0.80,0.71}
\definecolor{LightYellow4}{rgb}{0.55,0.55,0.48}
\definecolor{LightYellow}{rgb}{1.00,1.00,0.88}
\definecolor{LimeGreen}{rgb}{0.20,0.80,0.20}
\definecolor{MediumAquamarine}{rgb}{0.40,0.80,0.67}
\definecolor{MediumBlue}{rgb}{0.00,0.00,0.80}
\definecolor{MediumOrchid1}{rgb}{0.88,0.40,1.00}
\definecolor{MediumOrchid2}{rgb}{0.82,0.37,0.93}
\definecolor{MediumOrchid3}{rgb}{0.71,0.32,0.80}
\definecolor{MediumOrchid4}{rgb}{0.48,0.22,0.55}
\definecolor{MediumOrchid}{rgb}{0.73,0.33,0.83}
\definecolor{MediumPurple1}{rgb}{0.67,0.51,1.00}
\definecolor{MediumPurple2}{rgb}{0.62,0.47,0.93}
\definecolor{MediumPurple3}{rgb}{0.54,0.41,0.80}
\definecolor{MediumPurple4}{rgb}{0.36,0.28,0.55}
\definecolor{MediumPurple}{rgb}{0.58,0.44,0.86}
\definecolor{MediumSeaGreen}{rgb}{0.24,0.70,0.44}
\definecolor{MediumSlateBlue}{rgb}{0.48,0.41,0.93}
\definecolor{MediumSpringGreen}{rgb}{0.00,0.98,0.60}
\definecolor{MediumTurquoise}{rgb}{0.28,0.82,0.80}
\definecolor{MediumVioletRed}{rgb}{0.78,0.08,0.52}
\definecolor{MidnightBlue}{rgb}{0.10,0.10,0.44}
\definecolor{MintCream}{rgb}{0.96,1.00,0.98}
\definecolor{MistyRose1}{rgb}{1.00,0.89,0.88}
\definecolor{MistyRose2}{rgb}{0.93,0.84,0.82}
\definecolor{MistyRose3}{rgb}{0.80,0.72,0.71}
\definecolor{MistyRose4}{rgb}{0.55,0.49,0.48}
\definecolor{MistyRose}{rgb}{1.00,0.89,0.88}
\definecolor{NavajoWhite1}{rgb}{1.00,0.87,0.68}
\definecolor{NavajoWhite2}{rgb}{0.93,0.81,0.63}
\definecolor{NavajoWhite3}{rgb}{0.80,0.70,0.55}
\definecolor{NavajoWhite4}{rgb}{0.55,0.47,0.37}
\definecolor{NavajoWhite}{rgb}{1.00,0.87,0.68}
\definecolor{NavyBlue}{rgb}{0.00,0.00,0.50}
\definecolor{OldLace}{rgb}{0.99,0.96,0.90}
\definecolor{OliveDrab1}{rgb}{0.75,1.00,0.24}
\definecolor{OliveDrab2}{rgb}{0.70,0.93,0.23}
\definecolor{OliveDrab3}{rgb}{0.60,0.80,0.20}
\definecolor{OliveDrab4}{rgb}{0.41,0.55,0.13}
\definecolor{OliveDrab}{rgb}{0.42,0.56,0.14}
\definecolor{OrangeRed1}{rgb}{1.00,0.27,0.00}
\definecolor{OrangeRed2}{rgb}{0.93,0.25,0.00}
\definecolor{OrangeRed3}{rgb}{0.80,0.22,0.00}
\definecolor{OrangeRed4}{rgb}{0.55,0.15,0.00}
\definecolor{OrangeRed}{rgb}{1.00,0.27,0.00}
\definecolor{PaleGoldenrod}{rgb}{0.93,0.91,0.67}
\definecolor{PaleGreen1}{rgb}{0.60,1.00,0.60}
\definecolor{PaleGreen2}{rgb}{0.56,0.93,0.56}
\definecolor{PaleGreen3}{rgb}{0.49,0.80,0.49}
\definecolor{PaleGreen4}{rgb}{0.33,0.55,0.33}
\definecolor{PaleGreen}{rgb}{0.60,0.98,0.60}
\definecolor{PaleTurquoise1}{rgb}{0.73,1.00,1.00}
\definecolor{PaleTurquoise2}{rgb}{0.68,0.93,0.93}
\definecolor{PaleTurquoise3}{rgb}{0.59,0.80,0.80}
\definecolor{PaleTurquoise4}{rgb}{0.40,0.55,0.55}
\definecolor{PaleTurquoise}{rgb}{0.69,0.93,0.93}
\definecolor{PaleVioletRed1}{rgb}{1.00,0.51,0.67}
\definecolor{PaleVioletRed2}{rgb}{0.93,0.47,0.62}
\definecolor{PaleVioletRed3}{rgb}{0.80,0.41,0.54}
\definecolor{PaleVioletRed4}{rgb}{0.55,0.28,0.36}
\definecolor{PaleVioletRed}{rgb}{0.86,0.44,0.58}
\definecolor{PapayaWhip}{rgb}{1.00,0.94,0.84}
\definecolor{PeachPuff1}{rgb}{1.00,0.85,0.73}
\definecolor{PeachPuff2}{rgb}{0.93,0.80,0.68}
\definecolor{PeachPuff3}{rgb}{0.80,0.69,0.58}
\definecolor{PeachPuff4}{rgb}{0.55,0.47,0.40}
\definecolor{PeachPuff}{rgb}{1.00,0.85,0.73}
\definecolor{PowderBlue}{rgb}{0.69,0.88,0.90}
\definecolor{RosyBrown1}{rgb}{1.00,0.76,0.76}
\definecolor{RosyBrown2}{rgb}{0.93,0.71,0.71}
\definecolor{RosyBrown3}{rgb}{0.80,0.61,0.61}
\definecolor{RosyBrown4}{rgb}{0.55,0.41,0.41}
\definecolor{RosyBrown}{rgb}{0.74,0.56,0.56}
\definecolor{RoyalBlue1}{rgb}{0.28,0.46,1.00}
\definecolor{RoyalBlue2}{rgb}{0.26,0.43,0.93}
\definecolor{RoyalBlue3}{rgb}{0.23,0.37,0.80}
\definecolor{RoyalBlue4}{rgb}{0.15,0.25,0.55}
\definecolor{RoyalBlue}{rgb}{0.25,0.41,0.88}
\definecolor{SaddleBrown}{rgb}{0.55,0.27,0.07}
\definecolor{SandyBrown}{rgb}{0.96,0.64,0.38}
\definecolor{SeaGreen1}{rgb}{0.33,1.00,0.62}
\definecolor{SeaGreen2}{rgb}{0.31,0.93,0.58}
\definecolor{SeaGreen3}{rgb}{0.26,0.80,0.50}
\definecolor{SeaGreen4}{rgb}{0.18,0.55,0.34}
\definecolor{SeaGreen}{rgb}{0.18,0.55,0.34}
\definecolor{SkyBlue1}{rgb}{0.53,0.81,1.00}
\definecolor{SkyBlue2}{rgb}{0.49,0.75,0.93}
\definecolor{SkyBlue3}{rgb}{0.42,0.65,0.80}
\definecolor{SkyBlue4}{rgb}{0.29,0.44,0.55}
\definecolor{SkyBlue}{rgb}{0.53,0.81,0.92}
\definecolor{SlateBlue1}{rgb}{0.51,0.44,1.00}
\definecolor{SlateBlue2}{rgb}{0.48,0.40,0.93}
\definecolor{SlateBlue3}{rgb}{0.41,0.35,0.80}
\definecolor{SlateBlue4}{rgb}{0.28,0.24,0.55}
\definecolor{SlateBlue}{rgb}{0.42,0.35,0.80}
\definecolor{SlateGray1}{rgb}{0.78,0.89,1.00}
\definecolor{SlateGray2}{rgb}{0.73,0.83,0.93}
\definecolor{SlateGray3}{rgb}{0.62,0.71,0.80}
\definecolor{SlateGray4}{rgb}{0.42,0.48,0.55}
\definecolor{SlateGray}{rgb}{0.44,0.50,0.56}
\definecolor{SlateGrey}{rgb}{0.44,0.50,0.56}
\definecolor{SpringGreen1}{rgb}{0.00,1.00,0.50}
\definecolor{SpringGreen2}{rgb}{0.00,0.93,0.46}
\definecolor{SpringGreen3}{rgb}{0.00,0.80,0.40}
\definecolor{SpringGreen4}{rgb}{0.00,0.55,0.27}
\definecolor{SpringGreen}{rgb}{0.00,1.00,0.50}
\definecolor{SteelBlue1}{rgb}{0.39,0.72,1.00}
\definecolor{SteelBlue2}{rgb}{0.36,0.67,0.93}
\definecolor{SteelBlue3}{rgb}{0.31,0.58,0.80}
\definecolor{SteelBlue4}{rgb}{0.21,0.39,0.55}
\definecolor{SteelBlue}{rgb}{0.27,0.51,0.71}
\definecolor{VioletRed1}{rgb}{1.00,0.24,0.59}
\definecolor{VioletRed2}{rgb}{0.93,0.23,0.55}
\definecolor{VioletRed3}{rgb}{0.80,0.20,0.47}
\definecolor{VioletRed4}{rgb}{0.55,0.13,0.32}
\definecolor{VioletRed}{rgb}{0.82,0.13,0.56}
\definecolor{WhiteSmoke}{rgb}{0.96,0.96,0.96}
\definecolor{YellowGreen}{rgb}{0.60,0.80,0.20}
\definecolor{aliceblue}{rgb}{0.94,0.97,1.00}
\definecolor{antiquewhite}{rgb}{0.98,0.92,0.84}
\definecolor{aquamarine1}{rgb}{0.50,1.00,0.83}
\definecolor{aquamarine2}{rgb}{0.46,0.93,0.78}
\definecolor{aquamarine3}{rgb}{0.40,0.80,0.67}
\definecolor{aquamarine4}{rgb}{0.27,0.55,0.45}
\definecolor{aquamarine}{rgb}{0.50,1.00,0.83}
\definecolor{azure1}{rgb}{0.94,1.00,1.00}
\definecolor{azure2}{rgb}{0.88,0.93,0.93}
\definecolor{azure3}{rgb}{0.76,0.80,0.80}
\definecolor{azure4}{rgb}{0.51,0.55,0.55}
\definecolor{azure}{rgb}{0.94,1.00,1.00}
\definecolor{beige}{rgb}{0.96,0.96,0.86}
\definecolor{bisque1}{rgb}{1.00,0.89,0.77}
\definecolor{bisque2}{rgb}{0.93,0.84,0.72}
\definecolor{bisque3}{rgb}{0.80,0.72,0.62}
\definecolor{bisque4}{rgb}{0.55,0.49,0.42}
\definecolor{bisque}{rgb}{1.00,0.89,0.77}
\definecolor{black}{rgb}{0.00,0.00,0.00}
\definecolor{blanchedalmond}{rgb}{1.00,0.92,0.80}
\definecolor{blue1}{rgb}{0.00,0.00,1.00}
\definecolor{blue2}{rgb}{0.00,0.00,0.93}
\definecolor{blue3}{rgb}{0.00,0.00,0.80}
\definecolor{blue4}{rgb}{0.00,0.00,0.55}
\definecolor{blueviolet}{rgb}{0.54,0.17,0.89}
\definecolor{blue}{rgb}{0.00,0.00,1.00}
\definecolor{brown1}{rgb}{1.00,0.25,0.25}
\definecolor{brown2}{rgb}{0.93,0.23,0.23}
\definecolor{brown3}{rgb}{0.80,0.20,0.20}
\definecolor{brown4}{rgb}{0.55,0.14,0.14}
\definecolor{brown}{rgb}{0.65,0.16,0.16}
\definecolor{burlywood1}{rgb}{1.00,0.83,0.61}
\definecolor{burlywood2}{rgb}{0.93,0.77,0.57}
\definecolor{burlywood3}{rgb}{0.80,0.67,0.49}
\definecolor{burlywood4}{rgb}{0.55,0.45,0.33}
\definecolor{burlywood}{rgb}{0.87,0.72,0.53}
\definecolor{cadetblue}{rgb}{0.37,0.62,0.63}
\definecolor{chartreuse1}{rgb}{0.50,1.00,0.00}
\definecolor{chartreuse2}{rgb}{0.46,0.93,0.00}
\definecolor{chartreuse3}{rgb}{0.40,0.80,0.00}
\definecolor{chartreuse4}{rgb}{0.27,0.55,0.00}
\definecolor{chartreuse}{rgb}{0.50,1.00,0.00}
\definecolor{chocolate1}{rgb}{1.00,0.50,0.14}
\definecolor{chocolate2}{rgb}{0.93,0.46,0.13}
\definecolor{chocolate3}{rgb}{0.80,0.40,0.11}
\definecolor{chocolate4}{rgb}{0.55,0.27,0.07}
\definecolor{chocolate}{rgb}{0.82,0.41,0.12}
\definecolor{coral1}{rgb}{1.00,0.45,0.34}
\definecolor{coral2}{rgb}{0.93,0.42,0.31}
\definecolor{coral3}{rgb}{0.80,0.36,0.27}
\definecolor{coral4}{rgb}{0.55,0.24,0.18}
\definecolor{coral}{rgb}{1.00,0.50,0.31}
\definecolor{cornflowerblue}{rgb}{0.39,0.58,0.93}
\definecolor{cornsilk1}{rgb}{1.00,0.97,0.86}
\definecolor{cornsilk2}{rgb}{0.93,0.91,0.80}
\definecolor{cornsilk3}{rgb}{0.80,0.78,0.69}
\definecolor{cornsilk4}{rgb}{0.55,0.53,0.47}
\definecolor{cornsilk}{rgb}{1.00,0.97,0.86}
\definecolor{cyan1}{rgb}{0.00,1.00,1.00}
\definecolor{cyan2}{rgb}{0.00,0.93,0.93}
\definecolor{cyan3}{rgb}{0.00,0.80,0.80}
\definecolor{cyan4}{rgb}{0.00,0.55,0.55}
\definecolor{cyan}{rgb}{0.00,1.00,1.00}
\definecolor{darkblue}{rgb}{0.00,0.00,0.55}
\definecolor{darkcyan}{rgb}{0.00,0.55,0.55}
\definecolor{darkgoldenrod}{rgb}{0.72,0.53,0.04}
\definecolor{darkgray}{rgb}{0.66,0.66,0.66}
\definecolor{darkgreen}{rgb}{0.00,0.39,0.00}
\definecolor{darkgrey}{rgb}{0.66,0.66,0.66}
\definecolor{darkkhaki}{rgb}{0.74,0.72,0.42}
\definecolor{darkmagenta}{rgb}{0.55,0.00,0.55}
\definecolor{darkolive}{rgb}{0.33,0.42,0.18}
\definecolor{darkorange}{rgb}{1.00,0.55,0.00}
\definecolor{darkorchid}{rgb}{0.60,0.20,0.80}
\definecolor{darkred}{rgb}{0.55,0.00,0.00}
\definecolor{darksalmon}{rgb}{0.91,0.59,0.48}
\definecolor{darksea}{rgb}{0.56,0.74,0.56}
\definecolor{darkslate}{rgb}{0.18,0.31,0.31}
\definecolor{darkslate}{rgb}{0.18,0.31,0.31}
\definecolor{darkslate}{rgb}{0.28,0.24,0.55}
\definecolor{darkturquoise}{rgb}{0.00,0.81,0.82}
\definecolor{darkviolet}{rgb}{0.58,0.00,0.83}
\definecolor{deeppink}{rgb}{1.00,0.08,0.58}
\definecolor{deepsky}{rgb}{0.00,0.75,1.00}
\definecolor{dimgray}{rgb}{0.41,0.41,0.41}
\definecolor{dimgrey}{rgb}{0.41,0.41,0.41}
\definecolor{dodgerblue}{rgb}{0.12,0.56,1.00}
\definecolor{firebrick1}{rgb}{1.00,0.19,0.19}
\definecolor{firebrick2}{rgb}{0.93,0.17,0.17}
\definecolor{firebrick3}{rgb}{0.80,0.15,0.15}
\definecolor{firebrick4}{rgb}{0.55,0.10,0.10}
\definecolor{firebrick}{rgb}{0.70,0.13,0.13}
\definecolor{floralwhite}{rgb}{1.00,0.98,0.94}
\definecolor{forestgreen}{rgb}{0.13,0.55,0.13}
\definecolor{gainsboro}{rgb}{0.86,0.86,0.86}
\definecolor{ghostwhite}{rgb}{0.97,0.97,1.00}
\definecolor{gold1}{rgb}{1.00,0.84,0.00}
\definecolor{gold2}{rgb}{0.93,0.79,0.00}
\definecolor{gold3}{rgb}{0.80,0.68,0.00}
\definecolor{gold4}{rgb}{0.55,0.46,0.00}
\definecolor{goldenrod1}{rgb}{1.00,0.76,0.15}
\definecolor{goldenrod2}{rgb}{0.93,0.71,0.13}
\definecolor{goldenrod3}{rgb}{0.80,0.61,0.11}
\definecolor{goldenrod4}{rgb}{0.55,0.41,0.08}
\definecolor{goldenrod}{rgb}{0.85,0.65,0.13}
\definecolor{gold}{rgb}{1.00,0.84,0.00}
\definecolor{gray0}{rgb}{0.00,0.00,0.00}
\definecolor{gray100}{rgb}{1.00,1.00,1.00}
\definecolor{gray10}{rgb}{0.10,0.10,0.10}
\definecolor{gray11}{rgb}{0.11,0.11,0.11}
\definecolor{gray12}{rgb}{0.12,0.12,0.12}
\definecolor{gray13}{rgb}{0.13,0.13,0.13}
\definecolor{gray14}{rgb}{0.14,0.14,0.14}
\definecolor{gray15}{rgb}{0.15,0.15,0.15}
\definecolor{gray16}{rgb}{0.16,0.16,0.16}
\definecolor{gray17}{rgb}{0.17,0.17,0.17}
\definecolor{gray18}{rgb}{0.18,0.18,0.18}
\definecolor{gray19}{rgb}{0.19,0.19,0.19}
\definecolor{gray1}{rgb}{0.01,0.01,0.01}
\definecolor{gray20}{rgb}{0.20,0.20,0.20}
\definecolor{gray21}{rgb}{0.21,0.21,0.21}
\definecolor{gray22}{rgb}{0.22,0.22,0.22}
\definecolor{gray23}{rgb}{0.23,0.23,0.23}
\definecolor{gray24}{rgb}{0.24,0.24,0.24}
\definecolor{gray25}{rgb}{0.25,0.25,0.25}
\definecolor{gray26}{rgb}{0.26,0.26,0.26}
\definecolor{gray27}{rgb}{0.27,0.27,0.27}
\definecolor{gray28}{rgb}{0.28,0.28,0.28}
\definecolor{gray29}{rgb}{0.29,0.29,0.29}
\definecolor{gray2}{rgb}{0.02,0.02,0.02}
\definecolor{gray30}{rgb}{0.30,0.30,0.30}
\definecolor{gray31}{rgb}{0.31,0.31,0.31}
\definecolor{gray32}{rgb}{0.32,0.32,0.32}
\definecolor{gray33}{rgb}{0.33,0.33,0.33}
\definecolor{gray34}{rgb}{0.34,0.34,0.34}
\definecolor{gray35}{rgb}{0.35,0.35,0.35}
\definecolor{gray36}{rgb}{0.36,0.36,0.36}
\definecolor{gray37}{rgb}{0.37,0.37,0.37}
\definecolor{gray38}{rgb}{0.38,0.38,0.38}
\definecolor{gray39}{rgb}{0.39,0.39,0.39}
\definecolor{gray3}{rgb}{0.03,0.03,0.03}
\definecolor{gray40}{rgb}{0.40,0.40,0.40}
\definecolor{gray41}{rgb}{0.41,0.41,0.41}
\definecolor{gray42}{rgb}{0.42,0.42,0.42}
\definecolor{gray43}{rgb}{0.43,0.43,0.43}
\definecolor{gray44}{rgb}{0.44,0.44,0.44}
\definecolor{gray45}{rgb}{0.45,0.45,0.45}
\definecolor{gray46}{rgb}{0.46,0.46,0.46}
\definecolor{gray47}{rgb}{0.47,0.47,0.47}
\definecolor{gray48}{rgb}{0.48,0.48,0.48}
\definecolor{gray49}{rgb}{0.49,0.49,0.49}
\definecolor{gray4}{rgb}{0.04,0.04,0.04}
\definecolor{gray50}{rgb}{0.50,0.50,0.50}
\definecolor{gray51}{rgb}{0.51,0.51,0.51}
\definecolor{gray52}{rgb}{0.52,0.52,0.52}
\definecolor{gray53}{rgb}{0.53,0.53,0.53}
\definecolor{gray54}{rgb}{0.54,0.54,0.54}
\definecolor{gray55}{rgb}{0.55,0.55,0.55}
\definecolor{gray56}{rgb}{0.56,0.56,0.56}
\definecolor{gray57}{rgb}{0.57,0.57,0.57}
\definecolor{gray58}{rgb}{0.58,0.58,0.58}
\definecolor{gray59}{rgb}{0.59,0.59,0.59}
\definecolor{gray5}{rgb}{0.05,0.05,0.05}
\definecolor{gray60}{rgb}{0.60,0.60,0.60}
\definecolor{gray61}{rgb}{0.61,0.61,0.61}
\definecolor{gray62}{rgb}{0.62,0.62,0.62}
\definecolor{gray63}{rgb}{0.63,0.63,0.63}
\definecolor{gray64}{rgb}{0.64,0.64,0.64}
\definecolor{gray65}{rgb}{0.65,0.65,0.65}
\definecolor{gray66}{rgb}{0.66,0.66,0.66}
\definecolor{gray67}{rgb}{0.67,0.67,0.67}
\definecolor{gray68}{rgb}{0.68,0.68,0.68}
\definecolor{gray69}{rgb}{0.69,0.69,0.69}
\definecolor{gray6}{rgb}{0.06,0.06,0.06}
\definecolor{gray70}{rgb}{0.70,0.70,0.70}
\definecolor{gray71}{rgb}{0.71,0.71,0.71}
\definecolor{gray72}{rgb}{0.72,0.72,0.72}
\definecolor{gray73}{rgb}{0.73,0.73,0.73}
\definecolor{gray74}{rgb}{0.74,0.74,0.74}
\definecolor{gray75}{rgb}{0.75,0.75,0.75}
\definecolor{gray76}{rgb}{0.76,0.76,0.76}
\definecolor{gray77}{rgb}{0.77,0.77,0.77}
\definecolor{gray78}{rgb}{0.78,0.78,0.78}
\definecolor{gray79}{rgb}{0.79,0.79,0.79}
\definecolor{gray7}{rgb}{0.07,0.07,0.07}
\definecolor{gray80}{rgb}{0.80,0.80,0.80}
\definecolor{gray81}{rgb}{0.81,0.81,0.81}
\definecolor{gray82}{rgb}{0.82,0.82,0.82}
\definecolor{gray83}{rgb}{0.83,0.83,0.83}
\definecolor{gray84}{rgb}{0.84,0.84,0.84}
\definecolor{gray85}{rgb}{0.85,0.85,0.85}
\definecolor{gray86}{rgb}{0.86,0.86,0.86}
\definecolor{gray87}{rgb}{0.87,0.87,0.87}
\definecolor{gray88}{rgb}{0.88,0.88,0.88}
\definecolor{gray89}{rgb}{0.89,0.89,0.89}
\definecolor{gray8}{rgb}{0.08,0.08,0.08}
\definecolor{gray90}{rgb}{0.90,0.90,0.90}
\definecolor{gray91}{rgb}{0.91,0.91,0.91}
\definecolor{gray92}{rgb}{0.92,0.92,0.92}
\definecolor{gray93}{rgb}{0.93,0.93,0.93}
\definecolor{gray94}{rgb}{0.94,0.94,0.94}
\definecolor{gray95}{rgb}{0.95,0.95,0.95}
\definecolor{gray96}{rgb}{0.96,0.96,0.96}
\definecolor{gray97}{rgb}{0.97,0.97,0.97}
\definecolor{gray98}{rgb}{0.98,0.98,0.98}
\definecolor{gray99}{rgb}{0.99,0.99,0.99}
\definecolor{gray9}{rgb}{0.09,0.09,0.09}
\definecolor{gray}{rgb}{0.75,0.75,0.75}
\definecolor{green1}{rgb}{0.00,1.00,0.00}
\definecolor{green2}{rgb}{0.00,0.93,0.00}
\definecolor{green3}{rgb}{0.00,0.80,0.00}
\definecolor{green4}{rgb}{0.00,0.55,0.00}
\definecolor{greenyellow}{rgb}{0.68,1.00,0.18}
\definecolor{green}{rgb}{0.00,1.00,0.00}
\definecolor{grey0}{rgb}{0.00,0.00,0.00}
\definecolor{grey100}{rgb}{1.00,1.00,1.00}
\definecolor{grey10}{rgb}{0.10,0.10,0.10}
\definecolor{grey11}{rgb}{0.11,0.11,0.11}
\definecolor{grey12}{rgb}{0.12,0.12,0.12}
\definecolor{grey13}{rgb}{0.13,0.13,0.13}
\definecolor{grey14}{rgb}{0.14,0.14,0.14}
\definecolor{grey15}{rgb}{0.15,0.15,0.15}
\definecolor{grey16}{rgb}{0.16,0.16,0.16}
\definecolor{grey17}{rgb}{0.17,0.17,0.17}
\definecolor{grey18}{rgb}{0.18,0.18,0.18}
\definecolor{grey19}{rgb}{0.19,0.19,0.19}
\definecolor{grey1}{rgb}{0.01,0.01,0.01}
\definecolor{grey20}{rgb}{0.20,0.20,0.20}
\definecolor{grey21}{rgb}{0.21,0.21,0.21}
\definecolor{grey22}{rgb}{0.22,0.22,0.22}
\definecolor{grey23}{rgb}{0.23,0.23,0.23}
\definecolor{grey24}{rgb}{0.24,0.24,0.24}
\definecolor{grey25}{rgb}{0.25,0.25,0.25}
\definecolor{grey26}{rgb}{0.26,0.26,0.26}
\definecolor{grey27}{rgb}{0.27,0.27,0.27}
\definecolor{grey28}{rgb}{0.28,0.28,0.28}
\definecolor{grey29}{rgb}{0.29,0.29,0.29}
\definecolor{grey2}{rgb}{0.02,0.02,0.02}
\definecolor{grey30}{rgb}{0.30,0.30,0.30}
\definecolor{grey31}{rgb}{0.31,0.31,0.31}
\definecolor{grey32}{rgb}{0.32,0.32,0.32}
\definecolor{grey33}{rgb}{0.33,0.33,0.33}
\definecolor{grey34}{rgb}{0.34,0.34,0.34}
\definecolor{grey35}{rgb}{0.35,0.35,0.35}
\definecolor{grey36}{rgb}{0.36,0.36,0.36}
\definecolor{grey37}{rgb}{0.37,0.37,0.37}
\definecolor{grey38}{rgb}{0.38,0.38,0.38}
\definecolor{grey39}{rgb}{0.39,0.39,0.39}
\definecolor{grey3}{rgb}{0.03,0.03,0.03}
\definecolor{grey40}{rgb}{0.40,0.40,0.40}
\definecolor{grey41}{rgb}{0.41,0.41,0.41}
\definecolor{grey42}{rgb}{0.42,0.42,0.42}
\definecolor{grey43}{rgb}{0.43,0.43,0.43}
\definecolor{grey44}{rgb}{0.44,0.44,0.44}
\definecolor{grey45}{rgb}{0.45,0.45,0.45}
\definecolor{grey46}{rgb}{0.46,0.46,0.46}
\definecolor{grey47}{rgb}{0.47,0.47,0.47}
\definecolor{grey48}{rgb}{0.48,0.48,0.48}
\definecolor{grey49}{rgb}{0.49,0.49,0.49}
\definecolor{grey4}{rgb}{0.04,0.04,0.04}
\definecolor{grey50}{rgb}{0.50,0.50,0.50}
\definecolor{grey51}{rgb}{0.51,0.51,0.51}
\definecolor{grey52}{rgb}{0.52,0.52,0.52}
\definecolor{grey53}{rgb}{0.53,0.53,0.53}
\definecolor{grey54}{rgb}{0.54,0.54,0.54}
\definecolor{grey55}{rgb}{0.55,0.55,0.55}
\definecolor{grey56}{rgb}{0.56,0.56,0.56}
\definecolor{grey57}{rgb}{0.57,0.57,0.57}
\definecolor{grey58}{rgb}{0.58,0.58,0.58}
\definecolor{grey59}{rgb}{0.59,0.59,0.59}
\definecolor{grey5}{rgb}{0.05,0.05,0.05}
\definecolor{grey60}{rgb}{0.60,0.60,0.60}
\definecolor{grey61}{rgb}{0.61,0.61,0.61}
\definecolor{grey62}{rgb}{0.62,0.62,0.62}
\definecolor{grey63}{rgb}{0.63,0.63,0.63}
\definecolor{grey64}{rgb}{0.64,0.64,0.64}
\definecolor{grey65}{rgb}{0.65,0.65,0.65}
\definecolor{grey66}{rgb}{0.66,0.66,0.66}
\definecolor{grey67}{rgb}{0.67,0.67,0.67}
\definecolor{grey68}{rgb}{0.68,0.68,0.68}
\definecolor{grey69}{rgb}{0.69,0.69,0.69}
\definecolor{grey6}{rgb}{0.06,0.06,0.06}
\definecolor{grey70}{rgb}{0.70,0.70,0.70}
\definecolor{grey71}{rgb}{0.71,0.71,0.71}
\definecolor{grey72}{rgb}{0.72,0.72,0.72}
\definecolor{grey73}{rgb}{0.73,0.73,0.73}
\definecolor{grey74}{rgb}{0.74,0.74,0.74}
\definecolor{grey75}{rgb}{0.75,0.75,0.75}
\definecolor{grey76}{rgb}{0.76,0.76,0.76}
\definecolor{grey77}{rgb}{0.77,0.77,0.77}
\definecolor{grey78}{rgb}{0.78,0.78,0.78}
\definecolor{grey79}{rgb}{0.79,0.79,0.79}
\definecolor{grey7}{rgb}{0.07,0.07,0.07}
\definecolor{grey80}{rgb}{0.80,0.80,0.80}
\definecolor{grey81}{rgb}{0.81,0.81,0.81}
\definecolor{grey82}{rgb}{0.82,0.82,0.82}
\definecolor{grey83}{rgb}{0.83,0.83,0.83}
\definecolor{grey84}{rgb}{0.84,0.84,0.84}
\definecolor{grey85}{rgb}{0.85,0.85,0.85}
\definecolor{grey86}{rgb}{0.86,0.86,0.86}
\definecolor{grey87}{rgb}{0.87,0.87,0.87}
\definecolor{grey88}{rgb}{0.88,0.88,0.88}
\definecolor{grey89}{rgb}{0.89,0.89,0.89}
\definecolor{grey8}{rgb}{0.08,0.08,0.08}
\definecolor{grey90}{rgb}{0.90,0.90,0.90}
\definecolor{grey91}{rgb}{0.91,0.91,0.91}
\definecolor{grey92}{rgb}{0.92,0.92,0.92}
\definecolor{grey93}{rgb}{0.93,0.93,0.93}
\definecolor{grey94}{rgb}{0.94,0.94,0.94}
\definecolor{grey95}{rgb}{0.95,0.95,0.95}
\definecolor{grey96}{rgb}{0.96,0.96,0.96}
\definecolor{grey97}{rgb}{0.97,0.97,0.97}
\definecolor{grey98}{rgb}{0.98,0.98,0.98}
\definecolor{grey99}{rgb}{0.99,0.99,0.99}
\definecolor{grey9}{rgb}{0.09,0.09,0.09}
\definecolor{grey}{rgb}{0.75,0.75,0.75}
\definecolor{honeydew1}{rgb}{0.94,1.00,0.94}
\definecolor{honeydew2}{rgb}{0.88,0.93,0.88}
\definecolor{honeydew3}{rgb}{0.76,0.80,0.76}
\definecolor{honeydew4}{rgb}{0.51,0.55,0.51}
\definecolor{honeydew}{rgb}{0.94,1.00,0.94}
\definecolor{hotpink}{rgb}{1.00,0.41,0.71}
\definecolor{indianred}{rgb}{0.80,0.36,0.36}
\definecolor{ivory1}{rgb}{1.00,1.00,0.94}
\definecolor{ivory2}{rgb}{0.93,0.93,0.88}
\definecolor{ivory3}{rgb}{0.80,0.80,0.76}
\definecolor{ivory4}{rgb}{0.55,0.55,0.51}
\definecolor{ivory}{rgb}{1.00,1.00,0.94}
\definecolor{khaki1}{rgb}{1.00,0.96,0.56}
\definecolor{khaki2}{rgb}{0.93,0.90,0.52}
\definecolor{khaki3}{rgb}{0.80,0.78,0.45}
\definecolor{khaki4}{rgb}{0.55,0.53,0.31}
\definecolor{khaki}{rgb}{0.94,0.90,0.55}
\definecolor{lavenderblush}{rgb}{1.00,0.94,0.96}
\definecolor{lavender}{rgb}{0.90,0.90,0.98}
\definecolor{lawngreen}{rgb}{0.49,0.99,0.00}
\definecolor{lemonchiffon}{rgb}{1.00,0.98,0.80}
\definecolor{lightblue}{rgb}{0.68,0.85,0.90}
\definecolor{lightcoral}{rgb}{0.94,0.50,0.50}
\definecolor{lightcyan}{rgb}{0.88,1.00,1.00}
\definecolor{lightgoldenrod}{rgb}{0.93,0.87,0.51}
\definecolor{lightgoldenrod}{rgb}{0.98,0.98,0.82}
\definecolor{lightgray}{rgb}{0.83,0.83,0.83}
\definecolor{lightgreen}{rgb}{0.56,0.93,0.56}
\definecolor{lightgrey}{rgb}{0.83,0.83,0.83}
\definecolor{lightpink}{rgb}{1.00,0.71,0.76}
\definecolor{lightsalmon}{rgb}{1.00,0.63,0.48}
\definecolor{lightsea}{rgb}{0.13,0.70,0.67}
\definecolor{lightsky}{rgb}{0.53,0.81,0.98}
\definecolor{lightslate}{rgb}{0.47,0.53,0.60}
\definecolor{lightslate}{rgb}{0.47,0.53,0.60}
\definecolor{lightslate}{rgb}{0.52,0.44,1.00}
\definecolor{lightsteel}{rgb}{0.69,0.77,0.87}
\definecolor{lightyellow}{rgb}{1.00,1.00,0.88}
\definecolor{limegreen}{rgb}{0.20,0.80,0.20}
\definecolor{linen}{rgb}{0.98,0.94,0.90}
\definecolor{magenta1}{rgb}{1.00,0.00,1.00}
\definecolor{magenta2}{rgb}{0.93,0.00,0.93}
\definecolor{magenta3}{rgb}{0.80,0.00,0.80}
\definecolor{magenta4}{rgb}{0.55,0.00,0.55}
\definecolor{magenta}{rgb}{1.00,0.00,1.00}
\definecolor{maroon1}{rgb}{1.00,0.20,0.70}
\definecolor{maroon2}{rgb}{0.93,0.19,0.65}
\definecolor{maroon3}{rgb}{0.80,0.16,0.56}
\definecolor{maroon4}{rgb}{0.55,0.11,0.38}
\definecolor{maroon}{rgb}{0.69,0.19,0.38}
\definecolor{mediumaquamarine}{rgb}{0.40,0.80,0.67}
\definecolor{mediumblue}{rgb}{0.00,0.00,0.80}
\definecolor{mediumorchid}{rgb}{0.73,0.33,0.83}
\definecolor{mediumpurple}{rgb}{0.58,0.44,0.86}
\definecolor{mediumsea}{rgb}{0.24,0.70,0.44}
\definecolor{mediumslate}{rgb}{0.48,0.41,0.93}
\definecolor{mediumspring}{rgb}{0.00,0.98,0.60}
\definecolor{mediumturquoise}{rgb}{0.28,0.82,0.80}
\definecolor{mediumviolet}{rgb}{0.78,0.08,0.52}
\definecolor{midnightblue}{rgb}{0.10,0.10,0.44}
\definecolor{mintcream}{rgb}{0.96,1.00,0.98}
\definecolor{mistyrose}{rgb}{1.00,0.89,0.88}
\definecolor{moccasin}{rgb}{1.00,0.89,0.71}
\definecolor{navajowhite}{rgb}{1.00,0.87,0.68}
\definecolor{navyblue}{rgb}{0.00,0.00,0.50}
\definecolor{navy}{rgb}{0.00,0.00,0.50}
\definecolor{oldlace}{rgb}{0.99,0.96,0.90}
\definecolor{olivedrab}{rgb}{0.42,0.56,0.14}
\definecolor{orange1}{rgb}{1.00,0.65,0.00}
\definecolor{orange2}{rgb}{0.93,0.60,0.00}
\definecolor{orange3}{rgb}{0.80,0.52,0.00}
\definecolor{orange4}{rgb}{0.55,0.35,0.00}
\definecolor{orangered}{rgb}{1.00,0.27,0.00}
\definecolor{orange}{rgb}{1.00,0.65,0.00}
\definecolor{orchid1}{rgb}{1.00,0.51,0.98}
\definecolor{orchid2}{rgb}{0.93,0.48,0.91}
\definecolor{orchid3}{rgb}{0.80,0.41,0.79}
\definecolor{orchid4}{rgb}{0.55,0.28,0.54}
\definecolor{orchid}{rgb}{0.85,0.44,0.84}
\definecolor{palegoldenrod}{rgb}{0.93,0.91,0.67}
\definecolor{palegreen}{rgb}{0.60,0.98,0.60}
\definecolor{paleturquoise}{rgb}{0.69,0.93,0.93}
\definecolor{paleviolet}{rgb}{0.86,0.44,0.58}
\definecolor{papayawhip}{rgb}{1.00,0.94,0.84}
\definecolor{peachpuff}{rgb}{1.00,0.85,0.73}
\definecolor{peru}{rgb}{0.80,0.52,0.25}
\definecolor{pink1}{rgb}{1.00,0.71,0.77}
\definecolor{pink2}{rgb}{0.93,0.66,0.72}
\definecolor{pink3}{rgb}{0.80,0.57,0.62}
\definecolor{pink4}{rgb}{0.55,0.39,0.42}
\definecolor{pink}{rgb}{1.00,0.75,0.80}
\definecolor{plum1}{rgb}{1.00,0.73,1.00}
\definecolor{plum2}{rgb}{0.93,0.68,0.93}
\definecolor{plum3}{rgb}{0.80,0.59,0.80}
\definecolor{plum4}{rgb}{0.55,0.40,0.55}
\definecolor{plum}{rgb}{0.87,0.63,0.87}
\definecolor{powderblue}{rgb}{0.69,0.88,0.90}
\definecolor{purple1}{rgb}{0.61,0.19,1.00}
\definecolor{purple2}{rgb}{0.57,0.17,0.93}
\definecolor{purple3}{rgb}{0.49,0.15,0.80}
\definecolor{purple4}{rgb}{0.33,0.10,0.55}
\definecolor{purple}{rgb}{0.63,0.13,0.94}
\definecolor{red1}{rgb}{1.00,0.00,0.00}
\definecolor{red2}{rgb}{0.93,0.00,0.00}
\definecolor{red3}{rgb}{0.80,0.00,0.00}
\definecolor{red4}{rgb}{0.55,0.00,0.00}
\definecolor{red}{rgb}{1.00,0.00,0.00}
\definecolor{rosybrown}{rgb}{0.74,0.56,0.56}
\definecolor{royalblue}{rgb}{0.25,0.41,0.88}
\definecolor{saddlebrown}{rgb}{0.55,0.27,0.07}
\definecolor{salmon1}{rgb}{1.00,0.55,0.41}
\definecolor{salmon2}{rgb}{0.93,0.51,0.38}
\definecolor{salmon3}{rgb}{0.80,0.44,0.33}
\definecolor{salmon4}{rgb}{0.55,0.30,0.22}
\definecolor{salmon}{rgb}{0.98,0.50,0.45}
\definecolor{sandybrown}{rgb}{0.96,0.64,0.38}
\definecolor{seagreen}{rgb}{0.18,0.55,0.34}
\definecolor{seashell1}{rgb}{1.00,0.96,0.93}
\definecolor{seashell2}{rgb}{0.93,0.90,0.87}
\definecolor{seashell3}{rgb}{0.80,0.77,0.75}
\definecolor{seashell4}{rgb}{0.55,0.53,0.51}
\definecolor{seashell}{rgb}{1.00,0.96,0.93}
\definecolor{sienna1}{rgb}{1.00,0.51,0.28}
\definecolor{sienna2}{rgb}{0.93,0.47,0.26}
\definecolor{sienna3}{rgb}{0.80,0.41,0.22}
\definecolor{sienna4}{rgb}{0.55,0.28,0.15}
\definecolor{sienna}{rgb}{0.63,0.32,0.18}
\definecolor{skyblue}{rgb}{0.53,0.81,0.92}
\definecolor{slateblue}{rgb}{0.42,0.35,0.80}
\definecolor{slategray}{rgb}{0.44,0.50,0.56}
\definecolor{slategrey}{rgb}{0.44,0.50,0.56}
\definecolor{snow1}{rgb}{1.00,0.98,0.98}
\definecolor{snow2}{rgb}{0.93,0.91,0.91}
\definecolor{snow3}{rgb}{0.80,0.79,0.79}
\definecolor{snow4}{rgb}{0.55,0.54,0.54}
\definecolor{snow}{rgb}{1.00,0.98,0.98}
\definecolor{springgreen}{rgb}{0.00,1.00,0.50}
\definecolor{steelblue}{rgb}{0.27,0.51,0.71}
\definecolor{tan1}{rgb}{1.00,0.65,0.31}
\definecolor{tan2}{rgb}{0.93,0.60,0.29}
\definecolor{tan3}{rgb}{0.80,0.52,0.25}
\definecolor{tan4}{rgb}{0.55,0.35,0.17}
\definecolor{tan}{rgb}{0.82,0.71,0.55}
\definecolor{thistle1}{rgb}{1.00,0.88,1.00}
\definecolor{thistle2}{rgb}{0.93,0.82,0.93}
\definecolor{thistle3}{rgb}{0.80,0.71,0.80}
\definecolor{thistle4}{rgb}{0.55,0.48,0.55}
\definecolor{thistle}{rgb}{0.85,0.75,0.85}
\definecolor{tomato1}{rgb}{1.00,0.39,0.28}
\definecolor{tomato2}{rgb}{0.93,0.36,0.26}
\definecolor{tomato3}{rgb}{0.80,0.31,0.22}
\definecolor{tomato4}{rgb}{0.55,0.21,0.15}
\definecolor{tomato}{rgb}{1.00,0.39,0.28}
\definecolor{turquoise1}{rgb}{0.00,0.96,1.00}
\definecolor{turquoise2}{rgb}{0.00,0.90,0.93}
\definecolor{turquoise3}{rgb}{0.00,0.77,0.80}
\definecolor{turquoise4}{rgb}{0.00,0.53,0.55}
\definecolor{turquoise}{rgb}{0.25,0.88,0.82}
\definecolor{violetred}{rgb}{0.82,0.13,0.56}
\definecolor{violet}{rgb}{0.93,0.51,0.93}
\definecolor{wheat1}{rgb}{1.00,0.91,0.73}
\definecolor{wheat2}{rgb}{0.93,0.85,0.68}
\definecolor{wheat3}{rgb}{0.80,0.73,0.59}
\definecolor{wheat4}{rgb}{0.55,0.49,0.40}
\definecolor{wheat}{rgb}{0.96,0.87,0.70}
\definecolor{whitesmoke}{rgb}{0.96,0.96,0.96}
\definecolor{white}{rgb}{1.00,1.00,1.00}
\definecolor{yellow1}{rgb}{1.00,1.00,0.00}
\definecolor{yellow2}{rgb}{0.93,0.93,0.00}
\definecolor{yellow3}{rgb}{0.80,0.80,0.00}
\definecolor{yellow4}{rgb}{0.55,0.55,0.00}
\definecolor{yellowgreen}{rgb}{0.60,0.80,0.20}
\definecolor{yellow}{rgb}{1.00,1.00,0.00}

\renewcommand{\glat}{\bar g_{\rm lat}}

\begin{document}

\begin{titlepage}

\begin{flushright}
\small{
DESY 10-156 \\
SFB/CPP-11-51}
\end{flushright}

\begin{center}
{\Large\bf
One-loop lattice artifacts of a dynamical charm quark
\\
}
\end{center}
\vskip 0.35cm
\vbox{
\centerline{
\epsfxsize=2.8 true cm
\epsfbox{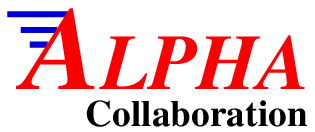}}
}
\vskip 0.1cm
\begin{center}
{Andreas Athenodorou and Rainer Sommer
}
\vskip 0.5cm
{
\vskip 2.0ex
NIC, DESY,
Platanenallee 6, 15738~Zeuthen,  Germany
\vskip 2.0ex
}
\vskip 0.5cm
{\bf Abstract}
\end{center}
\vskip 0.1ex
For a few observables in $\rmO(a)$ improved lattice QCD, we compute
discretization effects arising from the vacuum polarization 
of a heavy quark at one-loop order. In particular, the force between static quarks, 
the running coupling in the \SF
and a related quantity, $\vbar$, are considered. 
Results show that the cutoff effects
of a dynamical charm quark are typically smaller than those present in the 
pure gauge theory. This perturbative result is a good indication that 
dynamical charm quarks are feasible already now.


\vskip 2.0ex

\centerline{
September 2011
}

\vfill
\eject

\end{titlepage}

\section{Introduction}
Some collaborations are starting to include 
charm quark vacuum polarization in simulations
of lattice QCD \cite{2p1p1:milc1,2p1p1:etmc1,alpha:nf4,lat10:gregorio}.
Clearly the motivation is to exclude noticeable corrections due
to charm quark loops, especially when processes are considered
where intermediate to large momentum transfers are relevant. 
A particular such case is the extraction of the fundamental 
parameters of QCD, the $\Lambda$-parameter and quark masses.
A non-perturbative computation of these parameters 
needs a control of the theory 
over a large energy range, in the ideal case by a step scaling method
\cite{alpha:sigma,alpha:SU3,mbar:pap1,alpha:nf2,alpha:nf3}.
If such a computation is only based on simulations of the three-flavor theory,
the connection to the four and more flavor theory has to be done
perturbatively, without non-perturbative control. The 
ALPHA-collaboration has therefore started a computation in the 
four flavor theory\cite{alpha:nf4}. So far, only the running coupling in 
the massless \SF scheme has been considered. In order to set the
scale, i.e. determine the energy scale in $\GeV$, 
physical observables in the massive theory have to be 
computed in addition and matched to experiment. The question 
then arises whether this is 
feasible with the lattice spacings of around 
$0.1\,\fm \lesssim a \lesssim 0.05 \,\fm$ which are accessible
today. 

The main reason
to worry is that the charm quark mass in lattice units, $m_\charm a$,
is as large as 1/2 in this situation.\footnote{The
value of $m_\charm$ does of course depend on the renormalization scheme and 
the renormalization scale,
but for qualitative questions which we discuss here, this is not important.}
In the valence quark sector it has been found that up to such values
cutoff effects in the $\rmO(a)$-improved theory can be sizeable.
They do approximately show the expected form, quadratic in $a$ 
\cite{mbar:charm1,fds:JR03,fds:quenched2}, but it is also known
that above $am=1/2$ the Symanzik analysis of cutoff effects breaks
down and Symanzik $\rmO(a)$-improvement ceases to be 
useful \cite{zastat:pap2,hqet:pap2}. This knowledge has mainly 
been accumulated
in theories with a quenched charm quark; the large effects are due
to the propagation of a valence charm quark. A discussion of 
the cutoff effects of vacuum polarization contributions of heavy quarks
has been given in \Ref{pert:1loop}, again emphasizing the breakdown
of the Symanzik expansion.\footnote{Note 
in this context that the large effects found in \cite{2p1p1:etmc1}
are not of relevance in a theory with exact flavor symmetry. They can't
provide a guidance for the $\rmO(a)$-improved theory as defined in
\cite{impr:pap1}.}

In this paper we take a further step towards answering the question
of the size of cutoff effects in charm quark vacuum polarization effects.
We consider the leading such effects in perturbation theory,
namely we expand a few observables in the renormalized coupling and
examine the dependence of the first non-trivial 
expansion coefficient on the lattice 
spacing and the mass of the quark. 
We make use of our previous work 
with H. Panagopoulos\cite{pertforce:andreasharis}, extracting the 
cutoff effects in the force between static quarks and illustrate 
further the cutoff effects present in 
the fermionic contribution to the  \SF coupling\cite{pert:1loop}. 
Obviously the size of these effects is non-universal,
depending on all details of the regularization. Precise statements refer 
to the $\rmO(a)$ improved theory that we use. However, this does provide  
an example how much cutoff effects grow or do not with the 
mass of a dynamical quark. A shortcoming of our investigation
is that we just use perturbation theory. However concerning the qualitative picture
we do not think this restriction is too severe: the dominant 
contributions of charm quark loops is to processes with typical momenta
of the size of the charm mass. While such momenta are not large enough to 
expect high precision from
perturbation theory, we expect the qualitative features to 
be quite reliable.

\section{Lattice formulation, renormalization and cutoff effects
\label{s:lat}}

We start by recalling the most important features of the 
$\Oa$-improved theory which are relevant in the following.
Its definition is discussed in detail in \Ref{impr:pap1},
whose notation we use. Note that in principle the structure of the 
improved theory is more complicated when all quark masses are considered
\cite{impr:nondeg}, but this is irrelevant here, where we 
consider only one-loop effects.

In Wilson's regularization, the total action $S=S_g+S_f$ is given by
\begin{eqnarray}
  S_g[U] &=& \frac{1}{g_0^2}\sum_{p}w(p)\,\tr\,\{1-U(p)\} ,\label{e:Sg} \\
  S_f[U,\bar\psi,\psi] &=& 
     a^4\sum_{x}\sum_{i=1}^{\nf}\bar\psi_i(D+\mibare)\psi_i . \label{e:Sf}
\end{eqnarray}
The gauge field action $S_g$ is a sum over all oriented
plaquettes $p$ on the lattice, with the parallel transporters $U(p)$  around $p$.
The weights $w(p)$ are set to unity for now and we consider an infinite lattice. 
The Dirac operator is 
\begin{equation}
  D = \frac12 \sum_{\mu=0}^3
  \{\gamma_\mu(\nabla_\mu^\ast+\nabla_\mu^{})- a\nabla_\mu^\ast\nabla_\mu^{}\}
 + \csw\,\frac{ia}{4}\sum_{\mu,\nu=0}^3
 \,\sigma_{\mu\nu}\hat F_{\mu\nu}, \label{e:Dlat}
\end{equation}
with $\nabla_\mu$ and $\nabla_\mu^{\ast}$ the forward and backward 
covariant derivatives respectively.
The improvement term involving the
lattice approximation of the field strength 
$\hat F_{\mu\nu}$ (see \cite{impr:pap1} for its precise definition) 
has a coefficient \cite{impr:SW} 
\bes
  \csw=1+\rmO\left(g_0^2\right)\,.
\ees

The discussion of cutoff effects only makes sense after a renormalization of 
the theory and the cutoff effects do depend on the renormalization conditions.
Here we are interested in perturbation theory and we first  
choose massless renormalization schemes with a renormalization scale $\mu$. 

\subsection{Massless renormalization schemes}
\label{s:maslrenorm}
At the required order,
the renormalized coupling  
and quark masses are given by
\bes
  \gbar^2(\mu)=\gtilde^2\zg \left(\gtilde^2,a\mu\right),
  \quad
  \mir(\mu)=\mqitilde + \rmO\left(g_0^2\right),
\ees
in terms of the improved bare coupling and improved quark mass
\bes
   \gtilde^2=g_0^2\left(1+\bg^{(1)}g_0^2 a\sum^{N_f}_{i=1} \mibare\right) + \rmO\left(g_0^6\right), 
   \quad
  \mqitilde=\mibare\left(1-\frac12a\mibare\right)\,,
\ees
where\cite{pert:1loop,impr:pap1} 
\bes
  \bg^{(1)} = 0.01200(2)\,.
\ees
We remark once more that at higher orders in perturbation
theory, the structure of renormalization and improvement is more 
complicated in the case of non-degenerate quark masses
\cite{impr:nondeg}. 

Often one considers observables whose perturbative
expansion in $\gbarMSbar$ is known in the continuum theory.
It is then natural to renormalize in
the $\msbar$ scheme.
For clarity this is done in two steps. 
First we introduce the coupling in the {\em lattice} minimal subtraction
scheme, defined by subtracting order by order in 
perturbation theory just a polynomial in $\log(a\mu)$ without a 
constant part. For our purposes this means
\begin{eqnarray}
  \glat^2(\mu)=\gtilde^2\zlat\left(\gtilde^2,a\mu\right),
  \quad \zlat\left(\gtilde^2,a\mu\right) = 1 -2b_0 \gtilde^2 \log(a\mu) + \rmO\left(\gtilde^4\right)\,,
\end{eqnarray}
with $b_0=\frac1{(4 \pi)^2} \left( \frac{11}{3} N_c   - \frac{2}{3} N_f \right)$ and $N_c$ the number of colors.

Let us now consider an observable\footnote{By observable we simply mean 
a quantity which is free of
divergences after coupling renormalization and quark mass renormalization.} 
which depends on a single length scale $r$ in addition to the 
masses $\mir$ and which has an expansion\footnote{In general one has
$
  \bar O = \bar O^{(0)}(\vecz,a/r) + \bar O^{(1)}(\vecz,a/r)\,g_0^2 + \ldots\,
$
for a single scale observable with a regular expansion in $g_0^2$
made dimensionless using an appropriate 
power of $r$. 
Assuming $\bar O^{(1)}$ to be non-zero,
one may then form $O =  (\bar O - \bar O^{(0)})/\bar O^{(1)}$, which 
has the interpretation of a renormalized coupling and the expansion
\eq{e:Oexpansion1}.
}
\bes
  O = g_0^2 + O^{(1)}(\vecz,a/r)\,g_0^4 + \ldots\,, 
  \label{e:Oexpansion1}
\ees
with $\vecz = (z_1,\ldots,z_{\nf})$ and $z_i=\mir\cdot r$.
The observables that we use below are exactly of this form.
After inserting 
$g_0^2=\glat^2 + [2b_0 \log(a\mu) - \bg^{(1)} a\sum^{N_f}_{i=1} \mibare]\,\glat^4 +\ldots$ 
the renormalized expression has a unique split 
\bes
   O = \tilde O_\mrm{cont}\left(r\mu,\vecz,\glat^2(\mu)\right)
       \left(1 + \tilde\delta_O(r\mu,\vecz,\glat^2(\mu),a/r)\right),
\ees
into a continuum
piece and a lattice artifact
\bes
   \tilde\delta_O\left(r\mu,\vecz,\glat^2(\mu),a/r\right) =  
   \tilde\delta_O^{(0)}(r\mu,\vecz,a/r) +
   \tilde\delta_O^{(1)}(r\mu,\vecz,a/r) \glat^2(\mu) + \ldots \,,
\ees
with $\tilde\delta_O(r\mu,\vecz,\glat^2(\mu),0)=0$.
We may now switch renormalization to
the $\msbar$ scheme via the finite scheme transformation 
(independent of the lattice spacing)
\bes
   \glat^2(\mu) &=& \gbarMSbar^2 (\mu)  
   - \frac{c_{1}^{\rm lat,\MSbar}}{4 \pi} \gbarMSbar^4 (\mu) 
   + \rmO \left( \gbarMSbar^6  (\mu)    \right),
\label{e:defc1}
 \\
 c_1^{\rm lat,\MSbar}&=&c_{1,g} + \nf c_{1,f}\,,\quad 
 \\ 
 c_{1,g} &=& -\frac{ \pi }{2 \nc} + 2.135730074078457(2) \nc \,,\quad
 c_{1,f} = -0.39574962(2) \,,
\ees
where the coefficients are known 
from \Refs{pert:DashenGross,pert:HaseHase,pert:2loopLW} and 
\Refs{pert:1loop,Bode:2001uz}.
We then end up with the expansion of the relative lattice
artifacts
\bes
 \delta_O\left(r\mu,\vecz,\gbarMSbar^2(\mu),a/r\right) &\equiv&
 \left.{O - O_\mrm{cont} \over O_\mrm{cont}}\right|_{\gbarMSbar,\mir}
 \\ &=&   
   \delta_O^{(1)}(r\mu,\vecz,a/r) \gbarMSbar^2(\mu) + \ldots \,.
 \label{e:arteexp}
\ees
Our notation for the left hand side specifies explicitly 
how the theory is renormalized: $\gbarMSbar(\mu)$ and $\mir$ are held 
fixed as the continuum limit is approached.

A possibly confusing point is that these lattice artifacts are intrinsically
perturbative, because the way the theory is renormalized (by minimal 
subtraction) has no non-perturbative extension. The artifacts $\delta_O^{(i)}$
therefore can  not be regarded as perturbative expansion coefficients
of artifacts which appear in a non-perturbative
solution of the theory by a Monte Carlo computation. 
However, combinations
of the above defined $\delta^{(i)}$ from different observables yield the 
expansion coefficients of true non-perturbative artifacts. 
Let us briefly illustrate this.

As an example we choose to renormalize the theory through 
the \SF coupling $\gbarSF(L)$~\cite{SF:LNWW}
for massless quarks\cite{pert:1loop}. Non-perturbatively this means 
that for each value of $a$ the bare coupling is chosen such
that $\gbarSF(L)$ is held fixed. We follow the literature and keep
as argument of $\gbarSF$ the length scale $L$, related to the
energy renormalization scale as $\mu=1/L$.
Since the \SF coupling is an observable, its perturbation theory 
in terms of $\gbarMSbar^2$ is  just as described above,
\bes
   \gbarSF^2(L) = 
\left[\gbarMSbar^2(\mu) + c_1^{\rm SF,\MSbar} \gbarMSbar^4(\mu) + \ldots\right]
\cdot
\left[1 + \delta_\mrm{SF}^{(1)}(a/L) \gbarMSbar^2(\mu) + \ldots\right]\,.
\ees
Combining this expansion with \eq{e:arteexp} we have 
\bes
\label{e:deltaSF}
\delta_{O_{\rm SF}} \left(r\mu,\vecz,\gbarSF^2(L),a/r \right)
 &\equiv&
\left.{O - O_\mrm{cont} \over O_\mrm{cont}}\right|_{\gbarSF,\mir}
 \\
 &=&
   \left[\delta_O^{(1)}(r\mu,\vecz,a/r)- \delta_\mrm{SF}^{(1)}(a/L) \right]\, 
   \gbarSF^2(L) + \ldots \,. \nonumber
\ees
The combination 
$
\delta_O^{(1)}(r\mu,\vecz,a/r)- \delta_\mrm{SF}^{(1)}(a/L)
$ 
does describe the asymptotic behavior of 
the non-perturbatively defined left hand side at small
coupling, i.e. small $r$ and $L$.
 
In this paper we are interested in the cutoff effects 
induced by heavy quark vacuum polarization. We take observables 
without valence quarks (``pure gauge observables''), 
split all cutoff effects into a gluonic one 
and those due to the different quarks,
\bes
  \delta_O^{(1)}(r\mu,\vecz,a/r) &=& \delta_O^{(1,g)}(r\mu,a/r)
  +  \sum^{N_f}_{i=1}\delta_O^{(1,f)}(r\mu,z_i,a/r)
\ees
and are interested in particular, whether 
$\delta_O^{(1,f)}(r\mu,z,a/r)$ is large when $z$ is large,
as it may be the case for a charm quark. Holding the mass-less
\SF coupling fixed in taking the continuum limit,
the relevant contribution to the cutoff effect is then
$$
  \delta_O^{(1,f)}(r\mu,z,a/r) - \delta_\mrm{SF}^{(1,f)}(a/L) \approx 
  \delta_O^{(1,f)}(r\mu,z,a/r)\,.
$$
Here we have used the fact that for the standard \SF coupling
and massless quarks the contribution $\delta_\mrm{SF}^{(1,f)}(a/L)$ is
{\em very} small\cite{pert:1loop}. In \sect{s:force} we will, therefore,
simply discuss $\delta_O^{(1,f)}(r\mu,z,a/r)$ choosing as an observable
the force between static quarks.

\subsection{Massive renormalization schemes}
\label{s:masvrenorm}
While a massless renormalization scheme is attractive since 
it keeps the renormalization group equations simple, 
it is not the most convenient choice for non-perturbative computations in
QCD with a charm quark. The charm quark's mass is larger than the 
typical QCD scale. It therefore has reduced vacuum polarization effects 
which is most efficiently implemented by renormalizing at the finite mass
of the charm quark. (Of course the computation of true mass-effects is most 
easily done in a massless scheme.)

A physical massive scheme is defined by picking a specific
observable $O=O_0$ with the expansion \eq{e:Oexpansion1} and defining
the coupling in the massive scheme
\bes
  \gbar^2_{m}(\mu,\vecmr) \equiv O_0 = g_0^2 + O_0^{(1)}(\vecz,a/r)\,g_0^4 + \ldots\,,
  \label{e:Oexpansion} 
\ees
with $\mu=1/r_0$.
In principle this has to be supplemented by a condition for $\vecmr$, but we do not
need that at the perturbative order considered here.

Straight forwardly, as in  \eq{e:deltaSF}, we then have for a different observable $O$
or the same one at a different length scale $r$,
\bes
\label{e:deltamassive}
\delta_{O_{\rm m}} \hspace{-1.0mm} \left( \hspace{-0.5mm} r/r_0,\vecz,\gbarm^2(\mu,\vecmr\hspace{-0.5mm}),a/r \hspace{-0.5mm} \right)
 & \hspace{-3.0mm}\equiv \hspace{-3.0mm}&
\left.{O - O_\mrm{cont} \over O_\mrm{cont}}\right|_{\gbarm,\mir}
 \\
 &\hspace{-3.0mm}= \hspace{-3.0mm}&
   \left[\hspace{-0.5mm} \delta_O^{(1)} \hspace{-0.5mm} ( \hspace{-0.5mm} r\mu,r\vecmr,a/r \hspace{-0.5mm} ) \hspace{-0.5mm} - \hspace{-0.5mm}
         \delta_{O_0}^{(1)} \hspace{-0.5mm} ( \hspace{-0.5mm} r\mu,r_0\vecmr,a/r_0 \hspace{-0.5mm})\hspace{-0.5mm} \right]\, \hspace{-1.0mm}
   \gbarm^2(\mu,\vecmr \hspace{-0.5mm}) \hspace{-0.5mm} + \hspace{-0.5mm} \ldots \,. \nonumber
\ees
The cutoff effects in the massive scheme are just given as a combination 
of the ones in the massless scheme.
\section{The force between static quarks\label{s:force}}
As a first observable we consider the force $F(r)$ between static quarks
at a distance $r$.
In the continuum it is
defined in terms of the static potential as $F(r) = {d \over d r} V(r)$.
We need it in the lattice regularization. 
Since the potential is normally defined in terms of a non-local object, 
the Wilson loop,
it is not obvious that $\Oa$ improvement holds.  But one may relate
the potential exactly to correlators
of local fields in the Heavy Quark Effective Theory \cite{pot:intermed}
and thus concludes that $\Oa$ improvement does hold for differences
of potentials \cite{pot:intermed}.  For a pedagogical description
see \cite{LH:rainer}. In order to have an $\Oa$ improved force one then just
has to use a proper definition of the derivative. The most natural choice is 
$F_\mrm{naive}(r_\mrm{naive}) = \frac1a \left[ V((r,0,0))-V((r-a,0,0)) \right]$ 
with $r_\mrm{naive}=r-\frac{a}{2}$, when one uses the on-axis
potential as we do here. 
However, it is much better to define the force such that it has no
cutoff effects at the lowest order in perturbation theory as in 
\eq{e:Oexpansion1}.
This may be achieved by a different choice of the 
argument\cite{pot:r0},
\bes
  \label{e:defri}
  F( \rI) = \frac1a \left[ V((r,0,0))-V((r-a,0,0)) \right] = g_0^2{\cf \over 4\pi \rI^2} 
  + \rmO(g_0^4)\,, 
\ees
with $\cf = (N_c^2-1)/2N_c$. The improved radius $\rI$ is computed from the tree-level 
potential
\bes
  \label{e:vtree}
  V_\mrm{tree} (r) = \frac1a \cf \int^{\pi}_{-\pi} 
 {\rmd^3 k \over (2 \pi)^3} 
 {2 \sin^2 \left( \frac{k_1}{2} \frac{r}{a}  \right)  \over
  4 \sum^{3}_{j=1} {\rm sin}^2 \left( \frac{k_j}{2}  \right)},
\ees
by $\rI^{-2} = 4 \pi {\cf}^{-1}\frac1a [V_\mrm{tree}((r,0,0))-V_\mrm{tree}((r-a,0,0))]$.
See \Ref{pot:intermed} for an efficient way of computing \eq{e:vtree}.
In the same reference, one can also see that non-perturbatively
in the pure SU(3) gauge theory,
the force defined in terms of
$\rI$ has much smaller lattice artifacts compared to $r_\mrm{naive}$. 
We will use \eq{e:defri} in all numerical results that we show below.
We have, however, looked at all quantities also for $r_\mrm{naive}$.
There are no changes which are worth reporting about. 

\begin{figure}[tb!]
\centerline{\scalebox{0.9}{\includegraphics[height=7cm]{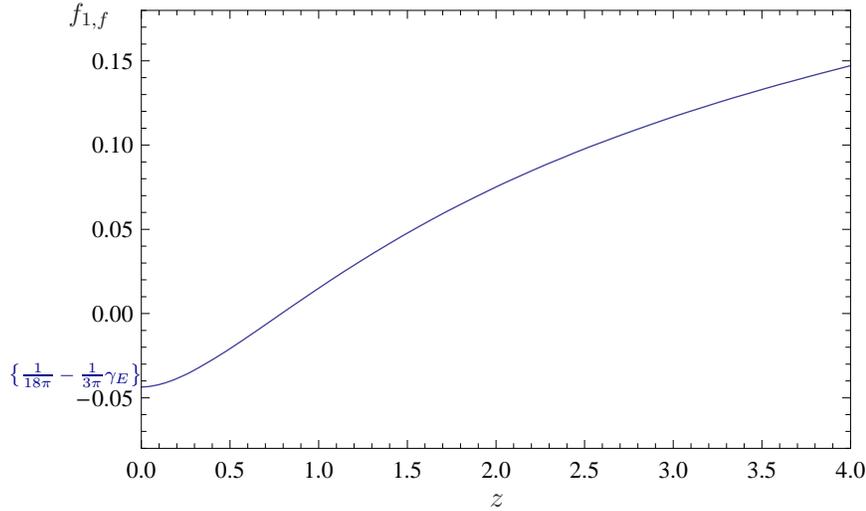} \put(-158,-10){\large $z$} \put(-335,194){$f_{1,f}$} \put(-360,43){{\scriptsize\color{DarkBlue}$\left\{  \frac{1}{18 \pi} - \frac{1}{3 \pi} \gamma_{E}  \right\}$}}}}
    \caption[]{\label{f:f1f} 
     The one-loop fermionic force contribution $f_{1,f} (z,0)$.}
\end{figure}

Loop corrections to the force are computed 
as described in \cite{pertforce:andreasharis}. We shall comment on details
below. For now we write down
immediately the expression renormalized
in the $\msbar$ scheme and set the renormalization scale
to the natural scale of the observable, $\mu=1/r$. 
\bes
  F = {\cf \alphaMSbar(1/r)  \over r^2} 
      \left\{1 + f_1(\vecz,a/r) \alphaMSbar(1/r) 
               + \rmO(\alphaMSbar^2) \right \}\,. 
\ees
From here on we have $r=\rI$ as the argument of the force
and quantities derived from it without indicating the subscript 
``I'' explicitly. 
The correction
term is split into gluonic and fermionic contributions
\bes
   f_1&=& f_{1,g}(a/r) + \sum^{N_f}_{i=1} f_{1,f} (z_i,a/r)\,.\quad 
\ees
For the determination of the lattice artifacts, we subtracted the
continuum expressions ($\gamma_E=0.57721566\dots$ is the Euler-Mascheroni constant)
\bes
   f_{1,g}(0) &=& {\nc \over \pi} \left[  - \frac{35}{36} +\frac{11}{6} \gamma_{E} \right], 
   \label{e:cng} \\
   f_{1,f} (z,0) & = & \frac{1}{2\pi} \left[ \frac{1}{3} {\rm log}(z^2)
                    + \frac{2}{3} \int_{1}^{\infty} \rmd x 
    \frac{1}{x^2} \sqrt{x^2-1}\left(1+\frac{1}{2x^2}  \right) 
                  \left( 1+2 z x  \right)e^{-2zx}   \right] \,,
   \label{e:cnf}
\ees 
which are derived from the ones for the 
potential\cite{pot:Fischler,pot:Billoire,pot:Melles,pot:Hoang}. 
The integral in \eq{e:cnf} is evaluated numerically. We show the fermion
contribution in \fig{f:f1f}.

The gluonic correction at finite $a/r$ is not our main concern.
For reference we will just cite
numbers from \cite{pot:pertBB}. 
However, we performed a re--evaluation of $f_{1,f}(z,a/r)$ following 
\Ref{pertforce:andreasharis}. 
For each value of $r/a$ the 
potential is given by a seven-dimensional integral over the Brillouin
zone. At fixed $a\mbare$ and
$r/a$ we evaluated these momentum integrals by
discretizing the Brillouin zone with a regular momentum lattice
with spacing  $\Delta k=2\pi / l$, applying the trapezoidal rule.
This procedure was carried out for
$r/a \in [1.358,15.467]$ and for more than thirty values of 
$a\mbare \in [-0.05,1.5]$. The force was then 
extrapolated in the momentum spacings, i.e. we
took the limit $l\to\infty$.
It is advantageous to extrapolate the force and not the potential
since the large unphysical self-energy contribution is then avoided.
The numerical results for the force at fixed $r/a$ were then
interpolated in $a\mbare$,
and the interpolations were subsequently used at the desired
values of $z$.
Interpolation and extrapolation errors form the uncertainties
visible in our figures below. 

Following our definition of the relative lattice artifacts, we have
\bes
  \left.{F - F_\mrm{cont} \over F_\mrm{cont}}\right|_{\gbarMSbar, \mir} &=&
  \delta_F^{(1)}(\vecz,a/r) \gbarMSbar^2(1/r), 
\\
  \delta_F^{(1)}(\vecz,a/r) &=& \delta_F^{(1,g)}(a/r) +
                            \sum_{i=1}^{\nf} \delta_F^{(1,f)}(z_i,a/r),
\ees
and, for example,
\bes
   4 \pi \delta_F^{(1,f)}(z,a/r) = f_{1,f} (z,a/r) - f_{1,f} (z,0)\,.
\ees
The factor $4\pi$ means that the 
relative size of the cutoff effect is given by multiplying with 
$\alpha=\gbar^2/(4\pi)$.

A first impression on the relevant overall
size of lattice artifacts is gained from the {\it gluonic} piece  $\delta_F^{(1,g)}(a/r)$.
The results of \Ref{pot:pertBB} are not precise enough to
see the asymptotic decay in $(a/r)^2$. Instead, we just extract 
\bes
  4\pi \delta_F^{(1,g)}(a/r)  &=& -0.232(\phantom{1}6) \text{ for } r/a=2.277\,,
\nonumber\\
\label{e:df1g}
  4\pi \delta_F^{(1,g)}(a/r) &=& -0.190(19) \text{ for } r/a=3.312\,,
\\
  4\pi\delta_F^{(1,g)}(a/r) &=& -0.151(42) \text{ for } r/a=4.319\,,
\nonumber
\ees
from the numbers listed in \Ref{pot:pertBB}. For the naive definition ($r_\mrm{naive}$)
of the lattice force, the cutoff effects are about a factor of
five larger and have the opposite sign compared to~(\ref{e:df1g}). Perturbation theory is probably a good 
guideline up to $\alpha=1/3$ at
which point we are then looking at around 5\% effects --
perturbation theory indicates that
larger values of $r/a$ are needed for precision physics. 

\begin{figure}[tb!]
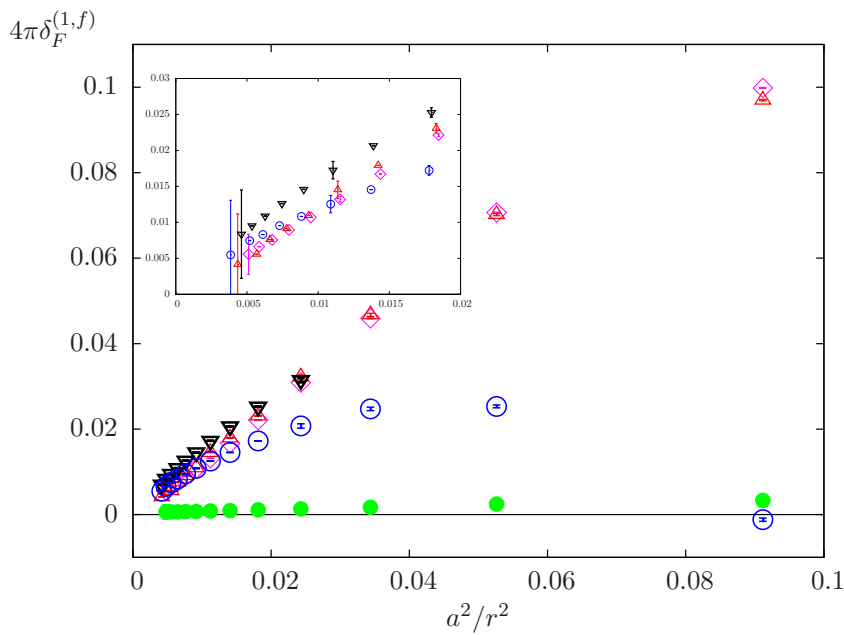

\centerline{\scalebox{0.93}{\input{plotwouterror.tex}}\put(-275,122){\scalebox{0.37}{\input{forcezoomnew2.tex}}}}
\vspace*{-3mm}
\caption[]{\label{f:df1f}
The fermionic cutoff effects $4\pi \delta_F^{(1,f)}(z,a/r)$.
For the improved theory ($\csw=1$, $\bg=0.01200(2)$) 
and our standard choice $\mir=\mqitilde$ we show
$z=0$~({\LARGE\color{magenta} $\diamond$}), 
$z=1$~({\small \color{red} $\triangle$}) and 
$z=3$~($\triangledown$) which is restricted to the range $a\mir < 1/2$. For the pole mass definition
we display just $z=3$~({\scriptsize\color{blue} $\bigcirc$}). Mass-less quarks with 
$\csw=\bg=0$ are given by {\LARGE\color{green} $\bullet$}.
We show some typical errors resulting from the $l$-extrapolation and the 
interpolation in $a m_0$.
}
\end{figure}

In \fig{f:df1f} we display some of our results for the {\it fermionic}
contribution to the cutoff effects $4\pi\delta_F^{(1,g)}$. Those following exactly
our description of $\Oa$-improvement ($\csw=1$) and renormalization ($\mir=\mqitilde$)
are shown by open symbols. The scale on the
y-axis is about a factor two smaller than the pure gauge effects \eq{e:df1g}. 
Generically the fermionic cutoff effects are smaller
than the gluonic ones and they {\em depend very little on the mass}. 
 
For illustration, we include in \fig{f:df1f} also a different
renormalization of the quark mass. In perturbation
theory, a possible one is
the ``pole mass'', $\mir = \frac1a\log(1+a\mibare) +\rmO(g_0^2)$.\footnote{We
note that there is a non-perturbative definition which coincides
with the pole mass at the lowest order of perturbation theory, 
which is relevant here. The non-perturbative mass is
$m_p = -{\rmd \over \rmd x_0} \log(\fp(x_0))$, with the \SF correlator
$\fp(x_0)$ as defined in \Ref{impr:pap1} for $\theta=0$ and
without background field.}
This mass can be taken larger in lattice units, which we do here only
to expose the resulting problematic cutoff effects. While these 
happen to be smaller than the cutoff effects for $\mir=\mqitilde$,
they have a rather non-linear behaviour as a function of $a^2/r^2$. 
We further remark that the asymptotic  $a^2/r^2$ scaling 
sets in later when $z$ is increased for this case.

A second remark concerns
mass-less quarks discretized with $\csw=0$. For mass-less quarks 
automatic $\Oa$ improvement holds in perturbation theory for
any value of $\csw$
\cite{tmqcd:FR1} 
(see \cite{nara:stefan} for a simple proof). The expected 
form of the cutoff effects is seen in the figure and their
magnitude happens to be smaller than for $\csw=1$.

Next, let us come back to the overall magnitude. 
Together with the factor $\alpha$ which accompanies
$4\pi \delta_F^{(1,f)}$, the magnitude is small. One has to 
remember that, since we consider a relative
cutoff effect, it is normalized to the leading order, 
$\nf$-independent, term. 
However, the whole fermionic contribution is relatively 
small. So in comparison to the physical effect of
the fermions, their cutoff effects are noticeable. 
Of course the continuum 
fermionic vacuum polarization contribution
depends on how one renormalizes the theory. In $\MSbar$-renormalization,
its smallness is just seen in \fig{f:f1f}, and in the non-perturbatively
defined SF-scheme (see the following section), one has 
an even smaller one-loop coefficient $f_{1,f}^\mrm{SF}$ in the relation 
$  F \hspace{-0.5mm}= \hspace{-0.5mm}{\cf \alphaSF(1/r)  \over r^2} \hspace{-0.5mm}
      \left\{ \hspace{-0.5mm} 1 + [f_{1,g}^\mrm{SF}+\nf f_{1,f}^\mrm{SF}] \alphaSF(1/r) 
               + \ldots \hspace{-0.5mm} \right \}\,
$ 
for massless quarks. Interestingly enough, the situation
is reversed when we consider the mass-dependence. The {\em physical} effect,
the variation in \fig{f:f1f}, is of the order of 0.2, while the 
{\em cutoff} effect is the {\em difference} of the data marked as triangles
and the massless ones marked as diamonds. These cutoff effects are tiny
compared to the physical ones. 

So far we have discussed the cutoff effects in a massless renormalization 
scheme. We may define a massive scheme by a coupling 
$\gbar^2_{m} = {4\pi r_0^2 \over \cf}F(r_0)$ and then
look at fixed $r_0\vecmr$ and $r/r_0$ how the continuum limit is approached.
Making use of \eq{e:deltamassive} the contribution\hspace{\stretch{1}} of\hspace{\stretch{1}}  a\hspace{\stretch{1}}
massive\hspace{\stretch{1}} quark\hspace{\stretch{1}} to\hspace{\stretch{1}} the\hspace{\stretch{1}} relative\hspace{\stretch{1}} cutoff\hspace{\stretch{1}} effects\hspace{\stretch{1}} in\hspace{\stretch{1}} $F(r)$\hspace{\stretch{1}} is\hspace{\stretch{1}} then \\
$\left[\delta_F^{(1,f)}(r \mr,a/r)- \delta_F^{(1,f)}(r_0\mr,a/r_0) \right]\, 
   \gbarm^2(1/r_0,\vecmr)+ \rmO(g^4)$.
Since $\delta_F^{(1,f)}(z,a/r)$ depends 
hardly on $z$, such cutoff effects are small as well as long as $a/r_0$ is small.

\section{The \SF coupling}

The \SF is the field theory in a finite space-time, taken here as $L^4$,
with Dirichlet boundary conditions in time and periodic boundary
condition (up to a phase $\theta$ for the quark fields) in 
space\cite{SF:LNWW,SF:stefan1,pert:1loop}.
The action can be written exactly as in \sect{s:lat} if the fields
outside the interval $0\leq x_0 \leq L$ are set to zero, the gauge links on
the boundaries $x_0=0$ and $x_0=L$ are set to fixed values 
and the fermion fields on the boundary are 
set to zero. 
Symanzik's effective action now contains terms 
localized at the boundaries
with new couplings. These result in cutoff effects linear
in the lattice spacing which are canceled by four more
improvement coefficients. One of them modifies the weight 
of the plaquettes touching the boundary to 
$w(p)= \ct=1-[0.08900 + 0.019141\nf \pm 0.00005]g_0^2 +\ldots$, 
the other ones play no r\^ole in our one-loop computation
due to our choice of the boundary conditions 
(abelian background field point ``A'' \cite{SF:LNWW})
and due to the order in
$g_0^2$. For more details we refer to \Refs{impr:pap1,SF:LNWW,pert:1loop}.

\begin{figure}[p!]
\centerline{\scalebox{0.93}{\input{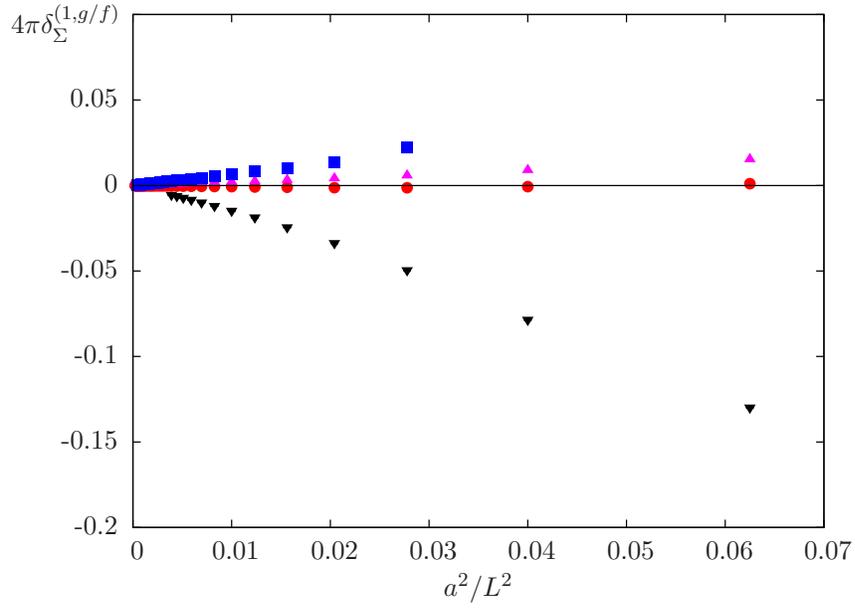}}}
\caption[]{\label{f:delsigpi5}
The cutoff effects $4\pi \delta_\Sigma^{(1,g)}(a/L)$ ({\large $\blacktriangledown$}) and  $4\pi  \delta_\Sigma^{(1,f)} (z,a/L)$ for $z=0$~({\LARGE\color{red} $\bullet$}), $z~=~1$~({\large\color{magenta} $\blacktriangle$}) and $z=3$~({\color{blue} $\blacksquare$}) extracted for $\theta=\pi/5$.
 }
\end{figure}
\begin{figure}[p!]
\centerline{\scalebox{0.93}{\input{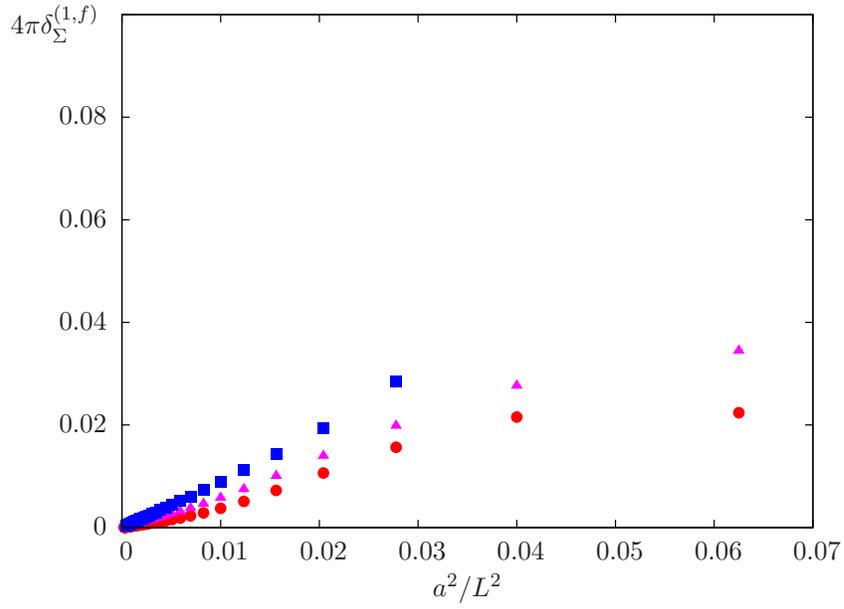}}}
\caption[]{\label{f:delsig0}
The cutoff effects 
$4\pi  \delta_\Sigma^{(1,f)} (z,a/L)$ for $z=0$~({\LARGE\color{red} $\bullet$}), $z=1$~({\large\color{magenta} $\blacktriangle$}) and $z=3$~({\color{blue} $\blacksquare$}) extracted for $\theta=0$.
}
\end{figure}

\begin{figure}[p!]
\centerline{\scalebox{0.93}{\input{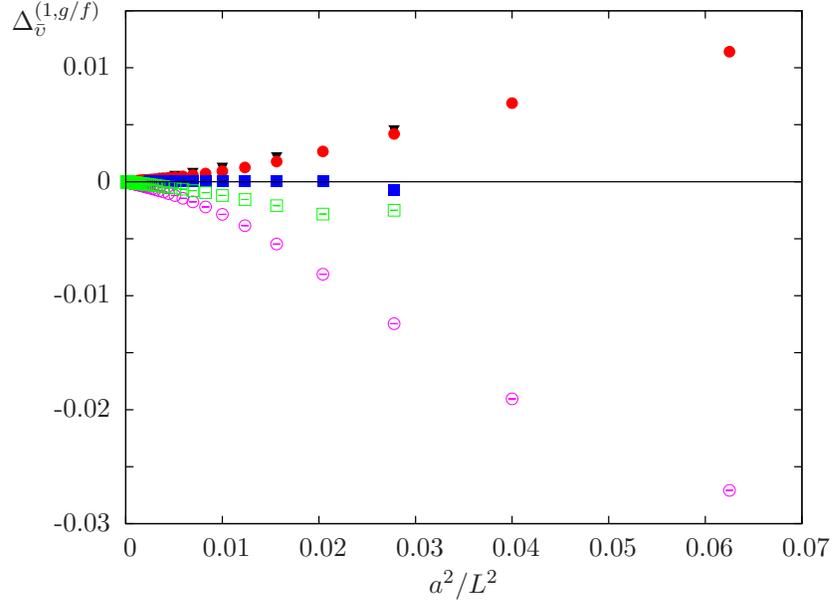}}}
\vspace*{-3mm}
\caption[]{\label{f:delvb0}
The cutoff effects 
$\Delta_{\vbar}^{(1,g)}(a/L)$ ({\small $\blacktriangledown$}) and $\Delta_{\vbar}^{(1,f)}(z,a/L)$  for $z=0$ ({\LARGE\color{red} $\bullet$}~({\scriptsize\color{magenta} $\bigcirc$})) and
$z=3$ ({\color{blue} $\blacksquare$} ({\color{green} $\square$})) extracted for $\theta=\pi/5 \ (\theta=0)$.
}
\end{figure}
\begin{figure}[p!]
\centerline{\scalebox{0.93}{\input{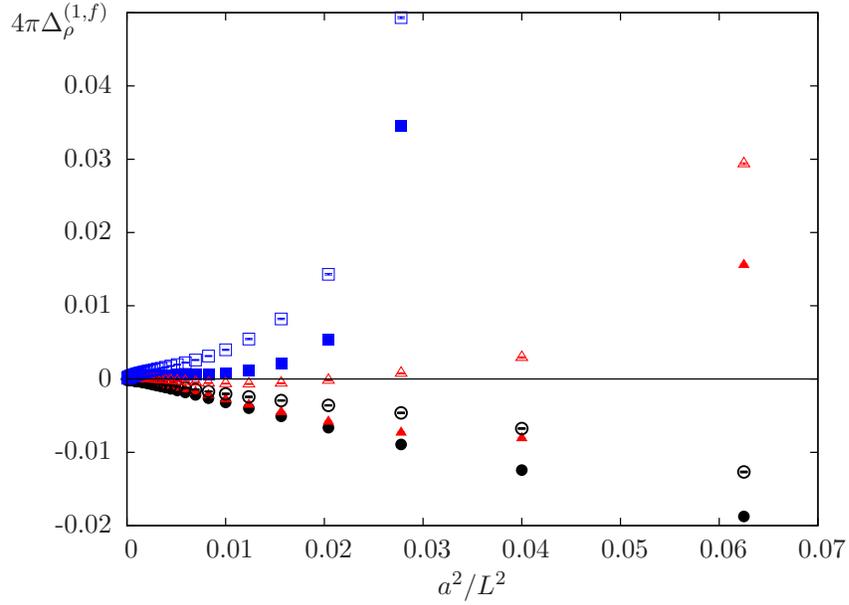}}}
\vspace*{-3mm}
\caption[]{\label{f:delrho}
The cutoff effects $4\pi  \Delta_{\rho}^{(1,f)}(z,a/L)$ for $z=1$~({\LARGE $\bullet$} ({\scriptsize $\bigcirc$})), $z=2$~({\large\color{red} $\blacktriangle$} ({\small\color{red} $\triangle$})) 
and $z=3$~({\color{blue} $\blacksquare$} ({\color{blue} $\square$})) extracted for $\theta=\pi/5$ ($\theta=0$).}
\end{figure}

The main virtue of the \SF is that it can be simulated for
mass-less quarks and a running coupling can be defined and computed
precisely in a Monte Carlo simulation (see \cite{nara:rainer} for a review).
We take the definition of the coupling from \Refs{SF:LNWW,pert:1loop}.
It depends on $L$ which plays the r\^ole of an inverse renormalization scale,
on the angle $\theta$ introduced above, and on a dimensionless parameter
$\nu$ which is usually set to zero. The dependence on $\nu$ is explicit,
\bes
  {1\over \gbar^2_\nu(L,\nu,\vecz)} = {1\over\gbarSF^2(L,\vecz)}-\nu\, \vbar(L,\vecz)\,.
\ees
We consider $\gbarSF$ and the quantity $\vbar$ which vanishes at
tree-level.  The mass-dependence is parameterized by the dimensionless
$z_i=\mir L$.

The central object needed in the non-perturbative computation of the running
of the coupling is the so-called step scaling function,
\bes
   \Sigma(u,\vecz,a/L) \equiv \gbarSF^2(2L,2z)|_{\gbarSF^2(L,\vecz)=u, \mir L=z_i}
    = u + \Sigma_1(\vecz,a/L)u^2 + \ldots \,,
\ees
but to see how cutoff effects behave we can also look at
other quantities, in particular
\bes
   \Omega(u,\vecz,a/L) \equiv \vbar(L,\vecz)|_{\gbarSF^2(L,\vecz)=u, \mir L=z_i} 
   = \vbar_1(\vecz,a/L)+\rmO(u)\,,
\ees
and 
\bes
   \rho(u,\vecz,a/L) \equiv 
   \frac{\gbarSF^2(L,\vecz) - \gbarSF^2(L,\veczero)}{\gbarSF^2(L,\veczero)} \biggr\vert_{\gbarSF^2(L,0)=u, \mir L=z_i} = 
   \rho_1(\vecz,a/L)u + \rmO(u^2)\,.
\ees
For the step scaling function $\Sigma$, we consider 
the relative lattice artifacts, defined as in \eq{e:deltaSF},
with the expansion 
\bes
   \delta_\Sigma(u,\vecz,a/L) = 
   {\Sigma(u,\vecz,a/L) - \Sigma(u,\vecz,0) \over \Sigma(u,\vecz,0)} 
   = \delta_\Sigma^{(1)}(\vecz,a/L) u + \ldots \,.
\ees
However, since both $\rho$ and $\vbar$ vanish at tree level it is
more convenient to consider the absolute artifacts,
\bes
   \Delta_{\vbar}(u,\vecz,a/L) &\equiv& \Omega(u,\vecz,a/L) - \Omega(u,\vecz,0)
   =
   \Delta_{\vbar}^{(1)}(\vecz,a/L)  + \rmO(u) \,,
\\
   \Delta_\rho(u,\vecz,a/L) &\equiv& \rho(u,\vecz,a/L) - \rho(u,\vecz,0)
   =
   \Delta_\rho^{(1)}(\vecz,a/L)u  + \rmO(u^2) \,.
\ees
Their decomposition into gluonic and fermionic pieces reads
\bes
 \delta_\Sigma^{(1)}(\vecz,a/L)  
   &=& \delta_\Sigma^{(1,g)}(\vecz,a/L) + \sum_{i=1}^{\nf}\delta_\Sigma^{(1,f)}(z_i,a/L),
   \\
 \Delta_{\vbar}^{(1)}(\vecz,a/L)  
   &=& \Delta_{\vbar}^{(1,g)}(a/L) + \sum_{i=1}^{\nf}\Delta_{\vbar}^{(1,f)}(z_i,a/L),
   \\
 \Delta_{\rho}^{(1)}(\vecz,a/L)  
   &=& \sum_{i=1}^{\nf}\Delta_\rho^{(1,f)}(z_i,a/L)\,.
\ees
 We note that $\Delta_{\rho}^{(1)}$
is a cutoff effect in a massless renormalization scheme (see \sect{s:maslrenorm}),
while  $\delta_\Sigma^{(1)}$,  $\Delta_{\vbar}^{(1)}$ refer to the massive 
\SF renormalization scheme (see \sect{s:masvrenorm}).

The artifact $\delta_\Sigma^{(1)}(\vecz,a/L)$ is illustrated 
in \fig{f:delsigpi5} for $\theta=\pi/5$ 
and for  $\theta=0$ in \fig{f:delsig0}.\footnote{The choice 
$\theta=\pi/5$ is the standard, since it was seen to be advantageous 
for Monte Carlo computations \cite{pert:1loop}, 
while $\theta=0$ is a natural alternative.} 
One observes that the cutoff effects for an individual $\Oa$ improved 
fermion are significantly smaller than the gluonic piece. 
For $\theta=\pi/5$, massless $\Oa$ improved 
fermions hardly show any cutoff effect -- a kinematical accident.
Generically cutoff effects grow with $z$, but not very much. 
Of course, one should not forget that for several fermion
flavors they add up accordingly. 

Compared to the step scaling function, the cutoff effects in $\vbar$,
\fig{f:delvb0},
are bigger, in particular when one takes into account that 
the relevant overall magnitude is $\vbar_1 \approx 0.1$ (that number
varies somewhat with the number of flavors and the parameter $\theta$).  
Massive quarks show smaller effects
than massless ones.

Finally, \fig{f:delrho} shows the cutoff effects in
$\rho$. These are to be compared to the continuum
values of $4\pi {\rho}(z,0) $ which range from 0.095 (0.086) 
for $z=1$ to 0.188 (0.170) for $z=3$
at $\theta=\pi/5$ ($\theta=0$).

\section{Conclusions}

In one-loop lattice perturbation theory, 
we investigated cutoff effects from including 
$\Oa$-improved Wilson quarks, and observed that in certain cases
they are comparable to the cutoff effects
in the pure gauge theory, see \fig{f:delvb0}. 
However, our main motivation for studying these 
was to check whether they become large
for massive quarks, in particular for values of
$am = 1/4 - 1/2$ which may be encountered in 
simulations with a charm quark. 

\begin{figure}[tb!]
\centerline{\scalebox{0.93}{\input{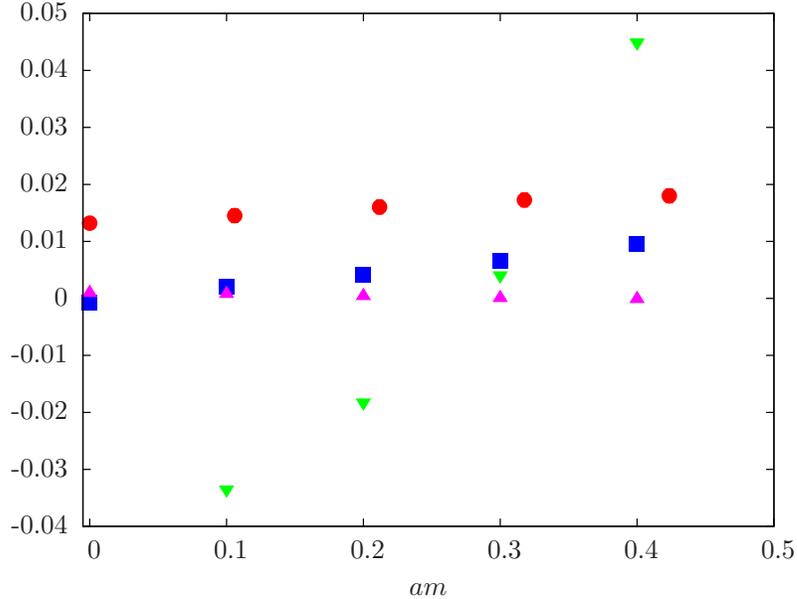}}}
\vspace*{-3mm}
\caption[]{\label{f:all}
Mass dependence of a few cutoff effects. 
We fix $a/r=1/9.445, a/L=1/10$ and show 
$4\pi \delta_F^{(1,f)}$ ({\LARGE\color{red} $\bullet$}) as well as
$4\pi \delta_\Sigma^{(1,f)}$ ({\color{blue} $\blacksquare$}) 
$\Delta_{\vbar}^{(1,f)}$ ({\large\color{magenta} $\blacktriangle$})
$\Delta_{\rho}^{(1,f)} /\rho^{(1,f)}(z,0) $ ({\large\color{green} $\blacktriangledown$}) for $\theta=\pi/5$ .
In contrast to our other graphs, the 
limit $am \to 0$ does not correspond to the continuum
limit since here we keep $a/L$ fixed and not $z$.
\\
}
\end{figure}

Contrary to expectation, we find that
the lattice artifacts do not grow
much and sometimes even become smaller for larger
masses. 
This statement is further illustrated in \fig{f:all}
where we choose a good enough resolution to have
small artifacts  for massless quarks and then display
them as a function of the mass in lattice units. 
They do remain small, which provides a good
indication that dynamical charm
quarks do not distort lattice QCD simulations at 
typical presently used lattice spacings.  


\vspace*{1cm}

\noindent
{\bf Acknowledgements.}

\noindent
We thank Haris Panagopoulos and Ulli Wolff for useful discussions,  
Hubert Simma for providing valuable comments on the manuscript and 
Mathias Steinhauser for pointing out \Ref{pot:Hoang}.
This work is supported by the Deutsche Forschungsgemeinschaft in the
SFB/TR 09. Computations were carried out
on the compute farm at DESY, Zeuthen.
\vspace*{1cm}

\bibliographystyle{JHEP}   
\bibliography{latticen,HQET}           

\end{document}